\documentclass[amsmath,amssymb,twocolumn,aps,jcp]{revtex4-2}

\usepackage{graphicx}
\usepackage{dcolumn}
\usepackage{bm}
\usepackage{hyperref}

\usepackage{amsmath}   
\usepackage{amsfonts}  
\usepackage{amssymb}   
\usepackage{graphicx}  

\begin{document}

\title{The Diffuse Layer of Thin Cylinders: Beyond the Counterion Condensation Paradigm}

\author{Olga I. Vinogradova}
\email[Corresponding author: ]{oivinograd@yahoo.com}
\affiliation{Frumkin Institute of Physical Chemistry and Electrochemistry, Russian Academy of Sciences, 31-4 Leninsky Prospect, 119071 Moscow, Russia}
\author{Elena F. Silkina}
\affiliation{Frumkin Institute of Physical Chemistry and Electrochemistry, Russian Academy of Sciences, 31-4 Leninsky Prospect, 119071 Moscow, Russia}

\date{\today}

\begin{abstract}
For over half a century, the counterion condensation theory has served as the dominant view for highly charged cylinders, prescribing a sharp charge renormalization based on the idealized abstraction of a salt-free line charge. In this work, we resolve this problem for physical open-domain systems across all charge regimes using the method of matched asymptotics applied to the non-linear Poisson--Boltzmann equation with added salt. For subcritical fields, where the dimensionless charge parameter satisfies $\xi < 1$, cylindrical curvature suppresses non-linear coupling, leaving the diffuse atmosphere unmodified. Beyond the critical threshold ($\xi = 1$), where the linear theory breaks down due to a localized Boltzmann explosion, the counterion accumulation smoothly transitions into a local flat-wall near-field geometry. By matching these zones, we construct a unified, parameter-free composite analytical solution capturing the spatial crossover toward the global cylindrical Debye screening tail. In the supercritical regime ($\xi > 1$), this near-field regularization locks the far-field effective charge into distinct, concentration-dependent parallel plateaus. Crucially, within the realistic regime where $\xi \le 5$, these idealized plateaus are not always reached, and the effective charge operates within a smooth, variable transition zone. Throughout all regimes, the ionic atmosphere remains diffuse, proving that the limiting laws of condensation emerge as artifacts of omitting the screening electrolyte.
\end{abstract}

\maketitle

\section{Introduction}

The electrostatic behavior of a charged cylinder immersed in an electrolyte solution is a fundamental problem in theoretical physics and colloid science, serving as a baseline model for electric double layers in open geometries. This geometry is of primary importance in polymer physics and biophysics, where it provides the theoretical archetype for rigid and semiflexible macromolecules, such as DNA, filamentous proteins, and synthetic polyelectrolytes. The distribution of electrostatic potential around the cylinder is governed by the second-order Poisson–Boltzmann equation. However, due to its non-linearity, closed-form analytical solutions exist only for a limited number of configurations. For a cylinder in an open domain, no uniform exact solution in elementary functions is available, leaving the literature reliant on either infinite series or numerical treatments. Historically, the search for a consistent description led to two distinct approaches, both designed to bypass the intrinsic logarithmic divergence of the salt-free cylindrical geometry.

The first approach, pioneered by \citet{fuoss.rm:1951} and \citet{alfrey.t:1951}, introduced the cylindrical cell model. By confining the charged cylinder within a concentric boundary of a finite radius $\mathcal{R}$, they imposed a localized electroneutrality condition by vanishing the electric field at the outer wall ($E(\mathcal{R}) = 0$). While this choice regularized the non-linear Poisson--Boltzmann equation, it restricted the solution to a dense packing of parallel cylinders where the fields of neighboring macroions cancel out at the cell interface. Consequently, this model prevented counterions from escaping the localized domain, tying the thermodynamic properties of an isolated macroion to the polymer concentration through the radius $\mathcal{R} \propto c^{-1/2}$~\cite{lebret1984distribution}. A comprehensive discussion can be found in the review by~\citet{andelman.d:2006}.

Seeking to isolate the physics of a single chain from such concentration-dependent geometry without abandoning the salt-free limit, Manning~\cite{manning.gs:1969} developed an alternative approach that provided a qualitative explanation for the low osmotic pressure observed in salt-free polyelectrolyte solutions. Rather than solving the full non-linear Poisson--Boltzmann equation, Manning analyzed the system through statistical mechanics, evaluating the configurational partition function of counterions surrounding an infinitely thin line charge. In this zero-radius continuum, the attractive energy of a counterion scales with the dimensionless charge parameter $\xi = \ell_B / b$ (where $\ell_B$ is the Bjerrum length and $b$ is the average charge spacing along the polyion axis) as $U(r) \propto 2\xi k_B T \ln r$, meaning its Boltzmann probability weight drops as $r^{-2\xi}$. When integrated over the two-dimensional volume element ($2\pi r dr$), the partition function scales as $\int r^{1-2\xi} dr$. Manning observed that for $\xi > 1$, this partition function diverges at the lower limit ($r \to 0$), signaling a thermodynamic instability where the electrostatic attraction overwhelms the translational entropy. To resolve this singularity, Manning postulated a transition: a fraction of counterions, $(\xi - 1)/\xi$, collapses onto the line charge to truncate its effective net charge down to the stability threshold, $\mathcal{A} = 1$, allowing the remaining ions to form a diffuse layer~\cite{manning.gs:1969}. Manning introduced this sharp two-phase partitioning primarily as a heuristic mathematical convenience; he explicitly noted that in real electrolyte environments with finite cylinder dimensions, the residual effective charge could float either above or below unity~\cite{manning.gs:1969}. However, after \citet{lebret1984distribution} demonstrated that this condensation behavior emerges from the non-linear Poisson--Boltzmann cell model in the absence of added salt and under the asymptotic limit of infinite dilution ($\mathcal{R} \to \infty$), subsequent literature rigidified this assumption into an invariant view of absolute charge locking. To trace the scaling physics, conformational transitions, and counterion condensation regimes in flexible charged macromolecules that emerged from this approach, we direct the reader to the review articles by \citet{barrat.jl:1996}, \citet{netz.review:2001}, \citet{dobrynin.av:2005}.

However, as recognized by~\citet{odijk.t:1977}, these limiting laws are idealized, proving valid only in the asymptotic limit of zero salt and zero radius. To incorporate salt screening, his subsequent description introduced a linear Yukawa potential. Namely, the cylindrical macromolecule was represented as a discrete chain of spherical beads---an approach that remains widespread also in salt-free analytical theories~\cite{gonzalez.p:1995} and coarse-grained molecular dynamics simulations~\cite{stevens.mj:1995}.

To resolve the non-physical singularities of unbounded electrostatics, subsequent descriptions sought a more consistent framework. As argued in the classic monograph by~\citet{grosberg.ayu:1994}, a rigorous treatment of the counterion accumulation requires solving the full non-linear Poisson--Boltzmann equation rather than relying on Manning's heuristic threshold. However, this mean-field path presents severe mathematical challenges. Subsequent frameworks reduced the cylindrical Poisson--Boltzmann equation to Painlev\'e III transcendents~\cite{tracy1997exact,tellez.g:2006}, but while these approaches offer formally exact asymptotic expansions, they result in non-closed, infinite series that remain cumbersome for calculations at finite salt concentrations. Alternatively, abandoning the mean-field approximation to explore strong-coupling correlation effects---such as the formation of correlated Wigner liquids near the surface~\cite{grosberg.ayu:2002,levin.y:2002}---proved physically prominent only for multivalent counterions. Consequently, the Poisson--Boltzmann equation is typically solved numerically, a task that presents its own computational challenges.

In attempting to address this mean-field challenge within the continuous Poisson--Boltzmann approach, several distinct theoretical efforts emerged. For instance, Vink sought to bridge continuous profiles with chemical concepts through explicit ion-binding formulations~\cite{vink.h:1990}. Among these, the seminal work established by \citet{fixman.m:1979} stands out as an electrostatic treatment that, unlike the idealized line-charge or bead-rod descriptions, preserved the continuous nature of the counterion atmosphere. This continuum perspective was further explored via asymptotic matching techniques by \citet{bocquet.l:2002}, who sought to evaluate how the non-linear equation regularizes the effective charge. However, for cylindrical geometries, such far-field matching frameworks inevitably reduce to the cumbersome Painlev\'e III transcendents, leaving the exact closed-form solution for the intermediate double layer out of reach. In parallel, instead of grappling with these intractable analytical representations, subsequent modeling was heavily drawn toward formalizing the qualitative image popularized by~\citet{grosberg.ayu:1994}, where counterions are conceptually confined within a tight accumulation ``layer'' around the cylinder. Attempting to translate this intuitive picture into a rigid calculational routine, both Fixman’s two-phase description~\cite{fixman.m:1979} and the subsequent numerical and simulation work by \citet{deserno.m:2000} engaged in a conceptual forcing. Seeking to replicate the sharp, salt-free asymptotics of the infinite-dilution cell limit established by \citet{lebret1984distribution}---a unique regime where the equation splits the ions into two distinct fractions---in salt-dominated environments, these authors introduced artificial geometric thresholds. Seeing that the presence of screening salt removes these sharp boundaries, leaving the exact non-linear profiles smooth, they relied on arbitrary definitions---such as finding the spatial minimum of $\rho(r)r^2$---to enforce a rigid ``condensation radius.'' All counterions within this arbitrary boundary were declared as bound mass to fit the historical $1 - 1/\xi$ rule~\cite{manning.gs:1969,deserno.m:2000}. The fallacy of this approach lies in its attempt to pin a far-field effective property onto a single, localized spatial coordinate within the non-linear double layer. Physically, the effective charge felt by the bulk electrolyte is not generated at a specific sharp boundary, but emerges from a continuous spatial integration across the diffuse cloud. By forcing this smooth, distributed transition to conform to an arbitrary, step-like geometric threshold, these works trapped the effective charge within a surface-centric description, leaving its salt-dependent far-field nature undetected.

This strategy of artificially removing mobile ions from the continuum descriptions reflects a broader historical pattern in physical chemistry. A century ago, \citet{stern.o:1924} introduced an immobile layer of adsorbed ions to describe the differential capacity of conductors, while \citet{bjerrum.n:1926} conceptualized the formation of localized ion pairs to interpret the electrical conductivity of salt solutions. In both fields, these concepts served the same purpose: they removed a portion of mobile ions from the continuum equations to fit experimental data, bypassing the missing physical mechanisms in the respective models. However, such assumptions are proving unnecessary. Recently, the conductivity of concentrated salt solutions was successfully described without invoking ion pairing~\cite{vinogradova.oi:2025}. Similarly, the large potential drop near surfaces, along with its effects on differential capacitance and surface forces, was resolved by introducing a diffuse Stern layer with a reduced solvent permittivity~\cite{uematsu.y:2018,vinogradova.oi:2024}.

For a cylinder with a finite radius surrounded by a background electrolyte, omitting the geometry forces literature to rely on adjustable parameters---such as empirical ion pairing and arbitrary effective charges---to fit experimental and computational data. This trend has shifted the focus toward multi-parameter numerical optimization rather than the development of physical descriptions. In this work, we show that this complexity is unnecessary. By applying the method of matched asymptotics to the non-linear cylindrical Poisson--Boltzmann equation with background salt, we construct a smooth, parameter-free composite analytical solution for the unbounded external domain. We prove that accounting for the geometric regularizations within the diffuse ion atmosphere captures the saturation behavior of the far-field effective charge. This approach eliminates the need for both the non-analytic charge-truncation postulates of the traditional Manning model and the artificial spatial cutoffs imposed by cell models. By showing that the counterions remain diffuse, our derivation renders the counterion condensation hypothesis redundant for moderate fields.

The remainder of the paper is arranged as follows. In Sec.~\ref{sec:framework}, the physical model is introduced and the exact integral relation between the surface charge and surface potential is derived. In Sec.~\ref{sec:analytics}, we examine the classical limits where the non-linear equation reduces to exactly solvable forms and derive a counterion-only solution in an open domain.
 Section~\ref{sec:perturbation} describes the matched asymptotic analysis with added salt. Section~\ref{sec:results} compares these analytical predictions with numerical solutions of the full non-linear equation. The limits of conventional theoretical approaches are discussed in Sec.~\ref{sec:paradigms}. We conclude in Sec.~\ref{sec:conclusions}.

\section{Governing Equations and Boundary Conditions}\label{sec:framework}

\begin{figure}[htbp]
    \centering
     \includegraphics[width=\columnwidth]{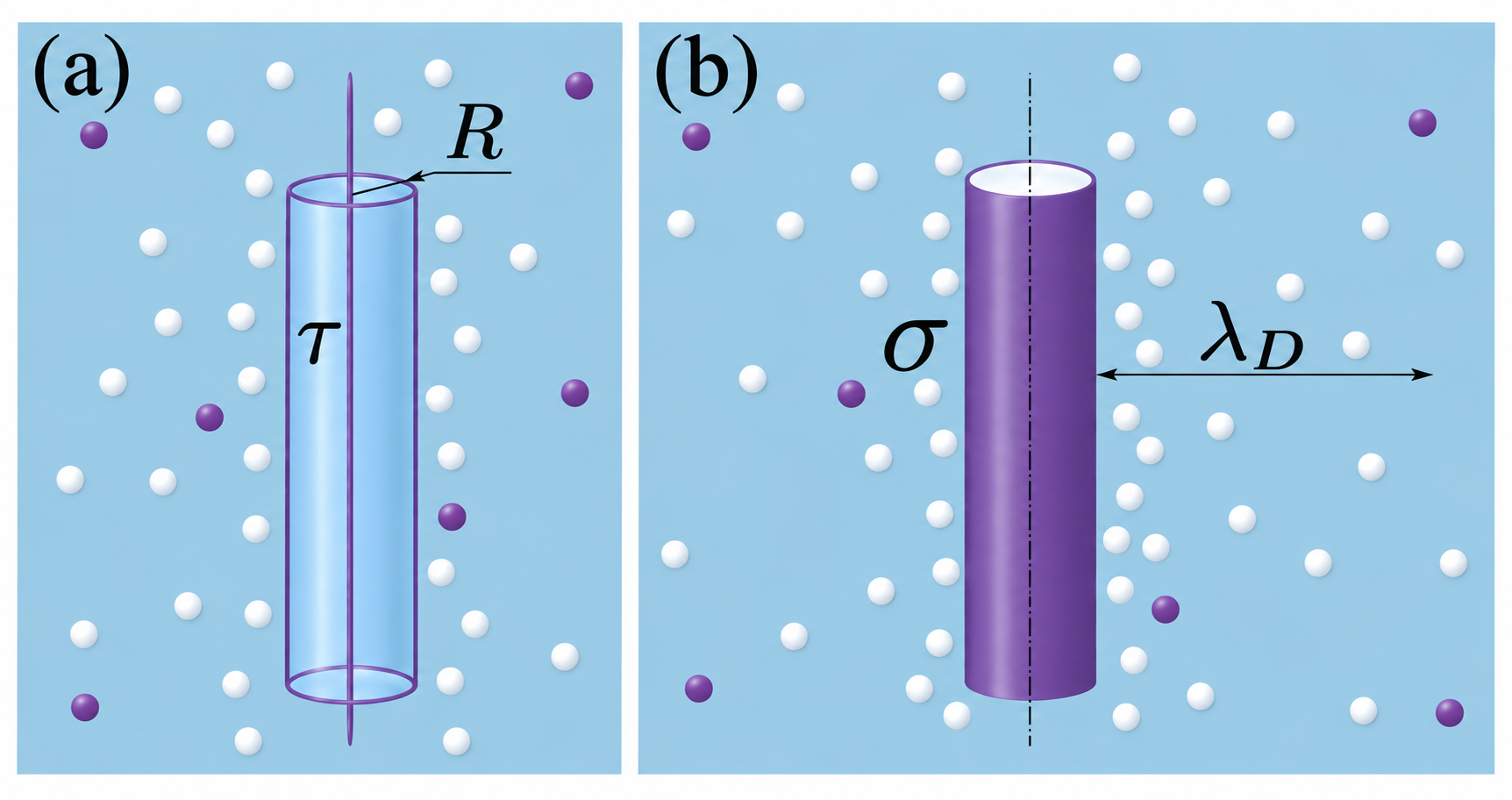}
    \caption{Schematic representation of the electrostatic models of a cylinder immersed in electrolyte solution.
    (a) The Manning model of an infinitely thin line with a line charge density $\tau$, but confined in a cylinder of radius $R$.
    (b) The present model of a cylinder with a surface charge density $\sigma = \tau / (2\pi R)$.
    Despite the different geometries, the parameter $\xi$ remains identical in both cases (see text).}
    \label{fig:manning_vs_cylinder}
\end{figure}

We consider a long nonconducting cylinder of radius $R$ and uniform surface charge density $\sigma$ immersed in a bulk symmetric 1:1 electrolyte solution with a dielectric permittivity $\epsilon$. According to the Gauss theorem, this configuration is electrostatically equivalent to an infinitely thin line charge of density $\tau = 2\pi R \sigma$ localized at the axis of a nonconducting cylinder of the same radius, as schematically illustrated in Fig.~\ref{fig:manning_vs_cylinder}. Mathematically, this representation maps directly onto the classical Manning model, with the crucial difference that the cylinder radius $R$ now introduces a finite excluded volume and the system includes an added salt. In response to the surface charge, an electrostatic diffuse layer (EDL) forms around the cylinder, with its thickness governed by the Debye screening length $\lambda_D$ of the bulk solution.

The electrostatic potential $\Phi(r)$ satisfies the non-linear cylindrical Poisson–Boltzmann equation:
\begin{equation}\label{eq:SPB}
\frac{1}{r} \frac{d}{d r} \left( r \frac{d \phi}{d r} \right) = \frac{d^2\phi}{dr^2} + \frac{1}{r}\frac{d\phi}{dr} = \lambda_D^{-2} \sinh\phi,
\end{equation}
where $\phi = \pm e\Phi/(k_B T)$ is the dimensionless potential, which is chosen to be inherently positive. To ensure this, the plus sign is taken for a positively charged cylinder, while the minus sign is used if the cylinder is negatively charged. Here, $e$ is the elementary charge, $k_{B}$ is the Boltzmann constant, and $T$ is the absolute temperature. The Debye screening length is defined as $\lambda_{D}=\left( 8\pi \ell _{B}n_{\infty}\right) ^{-1/2}$, where $n_\infty$ is the bulk ion number density, and $\ell _{B}=e^{2}/(\epsilon k_{B}T)$ denotes the Bjerrum length of the solvent. To satisfy global electroneutrality, the diffuse layer contains a total compensating charge of equal magnitude and opposite sign to that of the cylinder.

To solve Eq.~\eqref{eq:SPB}, we impose the far-field boundary conditions:
\begin{equation}\label{eq:infinity}
\phi(\infty) = 0, \quad \left.\frac{d\phi}{dr}\right|_{r=\infty} = 0,
\end{equation}
along with the Neumann boundary condition at the cylinder surface ($r=R$):
\begin{equation}\label{eq:CC}
\left.\frac{d\phi}{dr}\right|_{r=R} = -\frac{2}{\ell_{GC}},
\end{equation}
where $\ell_{GC} = \pm e/(2\pi \ell_B \sigma)$ denotes the Gouy–Chapman length, with the sign chosen to ensure that $\ell_{GC}$ remains positive.

Note that despite the different geometries—an infinitely thin line in the Manning model versus a cylinder of finite radius—the dimensionless parameter $\xi$ remains identical. In the Manning approach, it is defined as $\xi = \ell_B / b = \ell_B \tau / e$. In the cylinder model, it emerges as the ratio of the radius to the Gouy–Chapman length, $\xi = R / \ell_{GC}$. Substituting the definition of $\ell_{GC}$ yields exactly the same expression, establishing a direct physical link between the two models.

Multiplying both sides by $r$ and integrating from the cylinder surface $r=R$ to infinity ($r\to\infty$) yields, upon applying the boundary conditions~\eqref{eq:infinity}:
\begin{equation}
    -R \left.\frac{d\phi}{dr}\right|_{r=R} = \lambda_D^{-2} \int_R^\infty r \sinh\phi(r) \,dr.
    \label{eq:first_int_exact}
\end{equation}
Substituting the Neumann condition~\eqref{eq:CC} into Eq.~\eqref{eq:first_int_exact} and scaling by $\lambda_D / (2R)$ leads to the master integral relation:
\begin{equation}
    \frac{\lambda_D}{\ell_{\text{GC}}} = \frac{1}{2R\lambda_D} \int_R^\infty r \sinh\phi(r) \,dr.
    \label{eq:integral_master}
\end{equation}

Physically, Eq.~\eqref{eq:integral_master} expresses global electroneutrality. The integral on the right-hand side represents the total screening charge accumulated within the EDL, which balances the net surface charge of the cylinder represented by the left-hand side.
Although Eq.~\eqref{eq:integral_master} is exact, it cannot be solved in closed form because the integrand explicitly depends on the unknown potential profile $\phi(r)$. This relation highlights a defining mathematical feature of the problem: the cylinder radius $R$ enters both as a geometric prefactor and as the lower integration limit.

In the limit $R \to \infty$, Eq.~\eqref{eq:integral_master} must smoothly recover the classic planar Grahame relation:
\begin{equation} \label{eq:grahame}
\frac{\lambda_D}{\ell_{GC}} = \sinh\left(\frac{\phi_s}{2}\right).
\end{equation}

The behavior in the highly curved limit ($R/\lambda_D \to 0$) requires special comment. The electrostatic asymptotics near strongly charged geometries have traditionally been viewed as inherently non-linear, as a high surface charge density was expected to trigger severe exponential divergences within the full Poisson--Boltzmann framework. However, a closer inspection of Eq.~\eqref{eq:integral_master} reveals that the exact non-linear Poisson--Boltzmann equation automatically regulates this geometric coupling for cylindrical particles through a curvature-induced potential suppression. This can be rationalized as follows.

Because the potential profile $\phi(r)$ decays monotonically away from the surface, it is bounded from above by its surface value, $\phi(r) \le \phi_s$. Furthermore, since $\sinh\phi > \phi$ holds for any non-vanishing potential, the true non-linear screening charge in Eq.~\eqref{eq:integral_master} is strictly bounded from below by its linearized counterpart:
\begin{equation}\label{eq:app_ineq_cyl}
\frac{\lambda_D}{\ell_{GC}} > \frac{1}{2R\lambda_D} \int_R^\infty r \phi(r) \,dr.
\end{equation}
Under a fixed surface charge density ($\ell_{GC} = \text{const}$), the left-hand side of Eq.~\eqref{eq:app_ineq_cyl} remains invariant. While this strict inequality holds universally, the integral relation does not forbid it from becoming an approximate equality even for sufficiently large surface potentials. This means that the high cylindrical curvature can successfully suppress the non-linearity. However, this suppression is bounded, and at some point, the curvature can no longer prevent the non-linear growth.

To formalize this analysis, we define the fundamental small parameter of our asymptotic expansion as the ratio of the cylinder radius to the Debye length: $\varepsilon = R/\lambda_D \ll 1$. To resolve the stiff non-linearity, we split the domain into two distinct physical zones using appropriate inner and outer variables. Through this perturbation matching, we demonstrate below that the exact threshold where the curvature can no longer sustain this approximate equality is precisely $\xi = 1$.

\section{Exact Analytical Limits}\label{sec:analytics}

Before constructing the full dual-asymptotic framework, we examine the classical limits where the non-linear Poisson--Boltzmann equation reduces to exactly solvable forms. Rather than serving as mere reference points, these foundational limits expose the mathematical anomalies inherent to two-dimensional electrostatics, providing the strict boundary-layer asymptotes for our subsequent perturbation expansions. By analyzing the pure Laplacian field, the linearized Debye--Hückel framework, and the open-volume counter-ion-only solution in a unified manner, we map the exact boundaries where historical models break down. This systematic review uncovers how early theories were systematically forced to choose between the unphysical concentration-dependent constraints of closed cell frameworks and the heuristic, non-analytic postulates of linear condensation models, thereby establishing the precise mathematical vacuum that our self-consistent theory aims to fill.

\subsection{The Laplacian Limit (Pure Geometry)}	\label{subsec:laplacian}

In the total absence of mobile ions, the electrostatic behavior of the system is governed exactly by the Laplace equation in cylindrical coordinates:
\begin{equation}  \label{eq:L_cylinder}
\frac{d^2\phi}{dr^2} + \frac{1}{r}\frac{d\phi}{dr} = 0.
\end{equation}
Due to the intrinsic logarithmic divergence characteristic of two-dimensional electrostatics, the potential for an isolated charged cylinder in an infinite volume inevitably diverges toward minus infinity at large distances, $\phi(\infty) \to -\infty$, while the corresponding radial electric field naturally vanishes ($E \propto 1/r$).

Integrating Eq.~\eqref{eq:L_cylinder} and enforcing the Neumann boundary condition at the cylinder surface, $\phi'\vert_{r=R} = -2/\ell_{GC} = -2\xi/R$, yields the exact electrostatic potential profile:
\begin{equation}\label{eq:laplace_sol}
\phi(r) = \phi_s - 2\xi \ln\left(\frac{r}{R}\right),
\end{equation}
where $\phi_s \equiv \phi(R)$ represents the electrostatic potential at the surface.
 Crucially, within this purely geometric limit, $\phi_s$ cannot be determined uniquely and remains an arbitrary additive constant. Due to the lack of a well-defined reference point at infinity, this unattenuated logarithmic tail propagates infinitely into space—establishing a baseline divergence that any subsequent screening theory, such as the introduction of a finite cell radius $\mathcal{R}$ or mobile screening charges, must regularize.

\subsection{The Linearized Limit (Debye--Hückel Theory)}

If $\phi_s \le 1$, the approximation $\sinh\phi \simeq \phi$ holds, allowing Eq.~\eqref{eq:SPB} to be linearized into the Debye--Hückel framework:
\begin{equation}  \label{eq:DH_cylinder}
\frac{d^2\phi}{dr^2} + \frac{1}{r}\frac{d\phi}{dr} \simeq \lambda_D^{-2} \phi.
\end{equation}
The exact solution to Eq.~\eqref{eq:DH_cylinder} that satisfies the surface charge condition \eqref{eq:CC} and vanishes at infinity \eqref{eq:infinity} reads:
\begin{equation}\label{eq:DH_sol}
\phi(r) \simeq \frac{2 \lambda_D}{\ell_{GC}} \frac{K_0\left(\frac{r}{\lambda_D}\right)}{K_1\left(\frac{R}{\lambda_D}\right)},
\end{equation}
where $K_0$ and $K_1$ are the modified Bessel functions of the second kind of zeroth and first order, respectively.

In the limit of a thin cylinder or a highly dilute electrolyte background ($R/\lambda_D \ll 1$), Eq.~\eqref{eq:DH_sol} can be simplified significantly. Utilizing the leading-order asymptotic expansion for the denominator, $K_1(x) \simeq 1/x$, the full potential profile collapses into a compact form:
\begin{equation}
\phi(r) \simeq 2\xi K_0\left(\frac{r}{\lambda_D}\right). \label{eq:DH_expansion_thin}
\end{equation}

To examine the far-field behavior ($r \gg \lambda_D$), we invoke the long-range asymptotic expansion of the zeroth-order Bessel function, which yields:
\begin{equation}\label{eq:DH_far_field}
\phi(r) \simeq \sqrt{2\pi} \xi \sqrt{\frac{\lambda_D}{r}} e^{-r/\lambda_D}.
\end{equation}
Equation~\eqref{eq:DH_far_field} explicitly illustrates the exponential suppression that regularizes the system at large distances, ensuring that the far-field boundary condition is rigorously satisfied.

Conversely, the connection to the bare Laplacian framework becomes apparent in the near-field zone ($R \le r \ll \lambda_D$). By expanding Eq.~\eqref{eq:DH_expansion_thin} for small arguments, $K_0(x) \simeq \ln(2/x) - \gamma$, the potential profile near the cylinder takes the form:
\begin{equation}\label{eq:DH_near_field}
\phi(r) \simeq 2\xi \ln\left(\frac{2 e^{-\gamma} \lambda_D}{R}\right) - 2\xi \ln\left(\frac{r}{R}\right),
\end{equation}
where $\gamma \simeq 0.5772$ is the Euler--Mascheroni constant.
Evaluating Eq.~\eqref{eq:DH_near_field} directly at the cylinder surface ($r=R$) enables us to uniquely determine the surface potential $\phi_s$, a quantity that remained indefinite in the pure Laplace case:
\begin{equation}\label{eq:DH_sol2}
 \phi_s \simeq 2\xi \ln\left(\frac{2 e^{-\gamma} \lambda_D}{R}\right) \simeq 2\xi \ln\left(1.123\frac{\lambda_D}{R}\right).
\end{equation}
We remark and stress that in this limit $\phi_s$ is linear in $\xi$.

Substituting Eq.~\eqref{eq:DH_sol2} back into Eq.~\eqref{eq:DH_near_field} perfectly reproduces the Laplacian solution \eqref{eq:laplace_sol}. This cross-over behavior demonstrates that while the local spatial gradient near the surface retains its characteristic bare logarithmic scaling, the presence of the mobile electrolyte background regularizes the global electrostatics—resolving the mathematical gauge by establishing the scaled Debye length as a natural physical cutoff radius.

As a side note, this asymptotic structure exposes a flaw in contemporary molecular simulations of polyelectrolytes, where the polymer chain is modeled as a discrete string of beads interacting via a spherical Debye–Hückel (or Yukawa) potential, $\phi \propto e^{-r/\lambda_D}/r$. While a linear summation of such spherical fields reproduces the far-field cylindrical asymptote \eqref{eq:DH_far_field}, it introduces screening prematurely at the near-field scale. As shown by Eq.~\eqref{eq:DH_near_field}, the background electrolyte does not screen the cylinder at the single-bead level; instead, it acts as a global collective cutoff ($\lambda_D$) inside the logarithmic argument. By forcing individual spherical Debye screening onto each bead, Yukawa-based simulations linearize the near-field zone. This artificial regularization ignores the non-linear ion cloud around the charged cylinder, distorting the electrostatic potential and the ion distribution near the surface at higher charge densities where non-linear effects dominate.

\subsection{Counterions only in an infinite volume}
\label{subsec:counterion}

To establish a direct baseline for the proposed theory, we extend the counterion-only limit to a cylindrical geometry embedded in an infinite open volume ($r \in [R, \infty)$). Following the physical description established for the planar interface, we consider a long, positively charged cylinder characterized by a surface charge density $\sigma$. The negative counterions follow the Boltzmann distribution $n_{-}(r) = n_0 e^{\phi(r)}$, with the dimensionless potential field diverging at infinity, $\phi(\infty) \to -\infty$, which ensures $n_{-}(\infty) = 0$.

Under these geometric conditions, the non-linear Poisson--Boltzmann equation in cylindrical coordinates reads:
\begin{equation}
    \frac{d^2\phi}{dr^2} + \frac{1}{r}\frac{d\phi}{dr} = 4 \pi \ell_{B} n_0 e^{\phi}.
    \label{eq:PB_cyl}
\end{equation}
While the curvature term $\frac{1}{r}\frac{d\phi}{dr}$ prevents a first integration via multiplication by $\frac{d\phi}{dr}$ (as is standard in the planar case), despite a common belief Eq.~\eqref{eq:PB_cyl} admits an exact analytical solution given below.

Retaining the physical branch where $\phi \to -\infty$ as $r \to \infty$, the general solution for the electrostatic potential can be written by defining $r_0$ as the reference radius where the potential vanishes ($\phi(r_0) = 0$). By evaluating the Boltzmann relation at this reference position, $n_-(r_0) = n_0$, the integration constants are constrained by $2\pi \ell_{B} n_0 r_0^2 \beta^2 = 1$. This simplifies the potential profile to:
\begin{equation}
    \phi(r) = -2\ln\left[ \frac{r}{r_0} \frac{ \left(\frac{r}{r_0}\right)^\beta - \left(\frac{r}{r_0}\right)^{-\beta} }{2\beta} \right],
    \label{potential_cyl}
\end{equation}
where $\beta$ is the remaining independent scaling exponent. Substituting Eq.~\eqref{potential_cyl} into the Boltzmann factor yields the spatial concentration profile:
\begin{equation}
    n_{-}(r) = \frac{\beta^2}{2\pi \ell_{B} r^2 \sinh^2\left(\beta \ln \frac{r}{r_0}\right)}.
    \label{eq:cyl_conc}
\end{equation}
Equation~\eqref{eq:cyl_conc} demonstrates that the concentration distribution depends on the scaling exponent $\beta$ and the zero-potential radius $r_0$. In this unbounded geometry, these parameters are determined by matching the solution to the electrostatic Neumann boundary condition at the cylinder surface ($r=R$):
\begin{equation}
    \left.\frac{d\phi}{dr}\right|_{r=R} = -\frac{2}{R} \left[ 1 + \beta \coth\left(\beta \ln\frac{R}{r_0}\right) \right] = -\frac{2\xi}{R}.
    \label{eq:neumann_match}
\end{equation}
However, attempting to satisfy this boundary condition reveals a fundamental breakdown of the unbounded formulation itself.

Simplifying Eq.~\eqref{eq:neumann_match} yields an algebraic relation between the scaling exponent $\beta$ and the zero-potential radius $r_0$:
\begin{equation}
    \coth\left(\beta \ln\frac{R}{r_0}\right) = \frac{\xi - 1}{\beta}.
    \label{eq:coth_relation}
\end{equation}
In an open, salt-free environment, the global charge balance requires the total number of counterions in the infinite volume to neutralize the cylinder's surface charge, which dictates that $\beta = |1 - \xi|$. Substituting this condition into Eq.~\eqref{eq:coth_relation} reduces its right-hand side to $\text{sgn}(\xi - 1) = \pm 1$. However, because the hyperbolic cotangent function satisfies $\coth x = \pm 1$ only asymptotically as $x \to \pm \infty$, Eq.~\eqref{eq:coth_relation} cannot be satisfied for any finite reference radius $r_0$. This contradiction demonstrates the breakdown of the unbounded formulation, proving that a localized equilibrium profile cannot be sustained in an infinite open volume.

This divergence exposes the behavior of an isolated cylinder in an infinite, salt-free volume, splitting the system into two distinct regimes depending on the charge parameter $\xi$. For the subcritical case ($\xi < 1$), Eq.~\eqref{eq:coth_relation} demands that the zero-potential reference radius recedes to infinity ($r_0 \to \infty$), causing the counterion cloud to dilute. Conversely, for the supercritical regime ($\xi > 1$), the same relation forces $r_0$ into the interior of the cylinder ($r_0 < R$).

Under these conditions, attempts to verify global electroneutrality via the total charge integral become ill-posed:
\begin{equation}
    Q = \int_{R}^{\infty} 2\pi r n_{-}(r) \, dr.
    \label{eq:electroneutrality_cyl}
\end{equation}
The failure to establish a diffuse double layer manifests as two non-physical scenarios:
\begin{itemize}
    \item \textit{Subcritical regime ($\xi < 1$):} As $r_0 \to \infty$, the integration constant $\beta = 1 - \xi$ remains positive, but the global boundary condition forces the counterion cloud to expand indefinitely. The local density $n_-(r)$ vanishes at any finite distance, meaning the counterions undergo entropic evaporation, leaving the cylinder bare. Consequently, the integrand in Eq.~\eqref{eq:electroneutrality_cyl} collapses to zero.
    \item \textit{Supercritical regime ($\xi > 1$):} The exponent $\beta$ must change branches ($\beta = \xi - 1$), forcing the argument of the hyperbolic cotangent in Eq.~\eqref{eq:coth_relation} to flip sign. Since this shifts the zero-potential reference point inside the cylinder ($r_0 < R$), the total charge integral in Eq.~\eqref{eq:electroneutrality_cyl} diverges at the upper limit. To provide sufficient screening, the open system would require an infinite amount of counterion charge.
\end{itemize}
Thus, the open two-dimensional geometry is unstable at any charge density, as the total charge integral either vanishes or diverges at infinity.

Historically, to salvage the physics from this breakdown without abandoning the idealized salt-free limit, Manning introduced his classical condensation approach~\cite{manning.gs:1969}. Rather than questioning the unbounded geometry, he postulated a non-analytic transition: whenever the parameter $\xi$ exceeds unity, a fraction of counterions, $(\xi - 1)/\xi$, collapses onto the cylinder surface. This mechanism effectively truncates the net charge density down to $\xi = 1$, restoring the marginal convergence ($\beta = 0$) of the open-volume integral and preventing the ion atmosphere from blowing up.

As demonstrated in the subsequent sections, the necessity of invoking such sharp, heuristic charge modifications is merely an artifact of these historical approximations. By introducing a treatment of the non-linear diffuse layer in the presence of bulk salt, our approach bypasses the artificial charge-truncation postulates of open-space models. Crucially, this strategy also allows us to maintain an open-volume geometry without falling back on the rigid spatial boundaries imposed by traditional cell models.

\section{Matched Asymptotic Analysis with Added Salt}\label{sec:perturbation}

When the cylinder is much thinner than the Debye length ($R \ll \lambda_D$), a standard regular expansion fails to capture the coupling between the local boundary field and the global electrolyte screening cloud. In the salt-free limit, the cylindrical geometry introduces an inherent logarithmic divergence at large distances. Consequently, a straightforward regular perturbation series cannot remain uniformly valid throughout the entire unbounded volume. To rigorously untangle this boundary relation and capture how the ionic atmosphere regularizes the divergence in the high-curvature regime, we introduce the small parameter $\varepsilon = R/\lambda_D \ll 1$ and employ the method of matched asymptotics. Depending on the magnitude of the surface charge $\xi$, the leading-order governing equation within the inner zone reduces either to the Laplace equation in the subcritical regime ($\xi < 1$) or to the Liouville equation in the supercritical regime ($\xi > 1$).

\begin{figure}[h]
\centering
\includegraphics[width=0.45\textwidth]{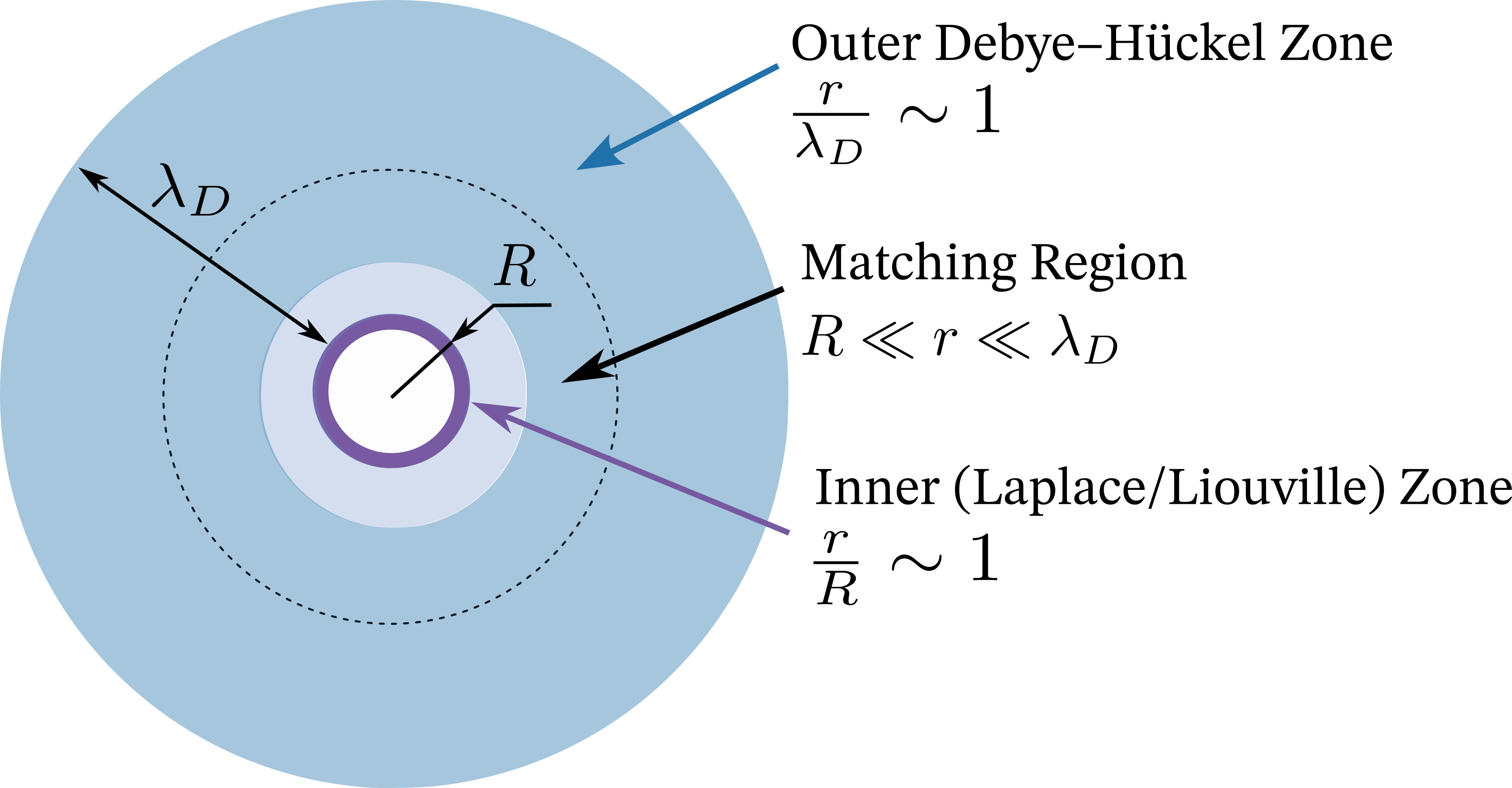}
\caption{Schematic decomposition of the diffuse layer around a charged cylinder in an open domain. The diagram highlights the inner zone ($r/R \sim 1$), where the full Poisson--Boltzmann equation asymptotically reduces to the Laplace ($\xi < 1$) or Liouville ($\xi > 1$) equation, and the outer zone ($\varepsilon(r/R) = r/\lambda_D \sim 1$) governed by the linear Debye--Hückel equation, which are smoothly coupled at the intermediate matching scale ($R \ll r \ll \lambda_D$).}
\label{fig:schematic}
\end{figure}

The system decomposes into two distinct regions governed by different physical length scales. The inner region is fixed to the cylinder surface, where the electrostatics is dominated by the geometric curvature. Here, the appropriate independent variable is the inner coordinate $\rho = r/R$, where $\rho = \mathcal{O}(1)$. In contrast, the outer region extends away from the cylinder, where the potential field is attenuated by the ionic atmosphere; the corresponding independent variable is the outer coordinate $\eta = r/\lambda_D$, with $\eta = \mathcal{O}(1)$. These scales are coupled via the spatial relation $\eta = \varepsilon \rho$.

Instead of forcing a fixed power series for the potentials, the asymptotic structures in both domains are determined self-consistently depending on the surface charge density $\xi$. To bridge the two spatial domains, we invoke the matching principle by evaluating the inner limit of the outer solution ($\eta \to 0$) and the outer limit of the inner solution ($\rho \to \infty$).

The potential remains sufficiently small in the far field to allow a linearization of the outer problem into the radial Debye–Hückel equation:
\begin{equation}\label{eq:outer_gov}
\frac{d^2\phi^{(o)}}{d\eta^2} + \frac{1}{\eta}\frac{d\phi^{(o)}}{d\eta} \simeq \phi^{(o)},
\end{equation}
yielding the Macdonald profile $\phi^{(o)}(\eta) \simeq M K_0(\eta)$, where $M$ is an integration constant. Expanding the modified Bessel function for small arguments and mapping the result onto the inner coordinate via the scaling relation $\eta = \varepsilon \rho$ recasts the outer solution into its local matching form:
\begin{equation}\label{eq:match_out_simplified}
\phi^{(o)} \simeq -M \ln\rho + M \ln\left( \frac{2}{\varepsilon} \right) - M \gamma,
\end{equation}
where subdominant terms of $\mathcal{O}(\varepsilon^2 \rho^2 \ln(\varepsilon\rho))$ are systematically omitted, and $\gamma$ is the Euler–Mascheroni constant.

In the inner zone, the electrostatic behavior depends fundamentally on the surface charge density $\xi$, leading to two distinct physical scenarios. In the subcritical regime ($\xi < 1$), the screening effects of the electrolyte become asymptotically negligible on the scale of the cylinder radius, causing the potential field to be governed by the Laplace equation. Conversely, in the supercritical regime ($\xi > 1$), strong counterion accumulation near the surface forces the inner problem to recover the geometry of the planar Poisson--Boltzmann equation, which reduces to the non-linear Liouville equation.

    Once the appropriate matching branch is evaluated, a unified, globally valid profile $\phi(r)$ can be constructed. Following the Prandtl matching protocol~\cite{Tollmien1961}, the composite expansion joins the near- and far-field solutions and subtracts their common overlap part to avoid double-counting:
\begin{equation}\label{eq:prandtl}
\phi(r) \simeq \phi^{(i)}(\rho) + \phi^{(o)}(\eta) - \left[\phi^{(i)}\right]^{(o)}.
\end{equation}

The global composite potential enables a direct calculation of the continuous spatial distributions for the mobile electrolyte ions throughout the entire diffuse layer. For a symmetric 1:1 background salt, the local number densities of the counter-ions ($n_-$) and co-ions ($n_+$) follow the Boltzmann distribution relative to the bulk concentration, $n_\pm(r) = n_\infty e^{\mp \phi(r)}$.

\subsection{Subcritical Regime}\label{subsec:mae}

\subsubsection{Asymptotic Matching Procedure}\label{subsec:mae1}

In the immediate vicinity of the cylinder ($\rho \sim 1$), the mobile ion screening term is heavily suppressed to $\mathcal{O}(\varepsilon^2)$. The leading-order inner potential answers strictly to the radial Laplace equation in terms of the high-curvature coordinate $\rho$:
\begin{equation}
\frac{d^2\phi^{(i)}}{d\rho^2} + \frac{1}{\rho}\frac{d\phi^{(i)}}{d\rho} \simeq 0.
\end{equation}
The exact general solution takes the classic logarithmic form:
\begin{equation}\label{eq:inner_gen}
\phi^{(i)}(\rho) \simeq -A \ln\rho + B,
\end{equation}
where $A$ is an integration constant and $B$ represents a reference potential that can implicitly depend on $\varepsilon$. Enforcing the electrostatic boundary condition at the cylinder interface ($\rho = 1$) immediately pins the first constant to the bare charge density, $A = 2\xi$. Consequently, the inner solution satisfies the surface charge condition exactly, while the remaining constant directly captures the surface potential, $\phi_s = \phi^{(i)}(1) = B$.

Crucially, dropping the screening term at this order is mathematically justified if and only if the exponential Boltzmann factor remains bounded, $\varepsilon^2 \sinh\phi_s \ll 1$. This condition reveals a striking physical threshold: the regular asymptotic matching scheme remains uniformly valid only within the subcritical charge regime $\xi < 1$. Beyond this point, the logarithmic growth of the near-field potential exponentially amplifies the Boltzmann term, instantly breaking the Laplacian dominance and forcing the non-linear electrolyte screening to lock directly onto the cylinder surface.

Enforcing the matching conditions between the outer solution in Eq.~\eqref{eq:match_out_simplified} and the inner solution in Eq.~\eqref{eq:inner_gen} within the intermediate overlap region requires both the coefficients of the $\ln\rho$ terms and the remaining constants to balance. This locks the integration constants:
\begin{equation}\label{eq:match_conditions}
M = 2\xi, \quad \text{and} \quad B = 2\xi \left[ \ln\left(\frac{2}{\varepsilon}\right) - \gamma \right].
\end{equation}
Substituting the expression for $B$ yields the surface potential:
\begin{equation}\label{HC_cylinder}
\phi_s \simeq 2\xi \left[ \ln\left( \frac{2}{\varepsilon} \right) - \gamma \right],
\end{equation}
which is equivalent to Eq.~\eqref{eq:DH_sol2}. This agreement confirms that for a thin cylinder, the matched asymptotic scheme recovers the high-curvature limit of the linear framework at the leading order. It underscores the nature of the background electrolyte in cylindrical geometries: the screening length $\lambda_D$ acts as the spatial cutoff that physically bounds the divergent line charge potential.

In contrast to the spherical geometry—where electrolyte screening acts as a subtractive correction to a finite Coulombic baseline~\cite{vinogradova.oi:2025}—Eq.~\eqref{HC_cylinder} reveals that for a cylinder, the screening length $\lambda_D$ enters inside the logarithm. The ionic atmosphere serves as the regulator that bounds the geometric divergence, making $\phi_s$ finite. Consequently, Eq.~\eqref{HC_cylinder} represents the high-curvature limit, establishing the analytical baseline from which the subsequent non-linear corrections unfold.

\subsubsection{Composite Global Solution and Ionic Concentrations}

Substituting the expressions and matching constraints into Eq.~\eqref{eq:prandtl}, and mapping all variables back to the radial distance $r$, the intermediate coordinate logarithmic terms cancel. This yields the global composite solution for the leading-order potential profile across the diffuse layer:
\begin{equation}
\phi(r) \simeq 2\xi K_0\left(\frac{r}{\lambda_D}\right). \label{eq:composite_final}
\end{equation}
Equation~\eqref{eq:composite_final} demonstrates that in this regime, the potential profile is governed by the charge parameter $\xi$ and the Debye length $\lambda_D$. Historically, theoretical descriptions of thin subcritical cylinders have employed linear approximations~\cite{fixman.m:1979}. The matched asymptotic derivation justifies this approach.

As established recently by \citet{vinogradova.sph:2026} for spherical particles, the geometric deactivation of near-field non-linearity does not require a low-potential constraint; instead, the linear Debye–Hückel framework remains asymptotically exact even for large surface potentials. For a sphere, this geometric constraint ensures that the linear potential matches the first term of the exact curvature expansion. For thin cylinders ($R/\lambda_D \ll 1$), this curvature-driven suppression of non-linear coupling causes the electrostatic profile in the subcritical regime to recover the Debye–Hückel form given by Eq.~\eqref{eq:DH_expansion_thin}.

To find the radial distribution of counterions in the near-zone ($r \to R$), we substitute the near-wall potential into the Boltzmann factor. This yields a power law for the density profile:
\begin{equation}
\frac{n_-(r)}{n_\infty} \simeq e^{\phi_s} \left( \frac{R}{r} \right)^{2\xi}.
\label{eq:conc_near_wall}
\end{equation}
Applying the boundary relation~\eqref{HC_cylinder}, the radius $R$ cancels out of the prefactor, relating the local counterion density to the bulk electrolyte scale:
\begin{equation}
\frac{n_-(\rho)}{n_\infty} \simeq \left( \frac{2 e^{-\gamma}}{\varepsilon} \right)^{2\xi} \left( \frac{1}{\rho} \right)^{2\xi}  = \left( \frac{2 e^{-\gamma} \lambda_D}{r} \right)^{2\xi}.
\label{eq:conc_near_wall_explicit}
\end{equation}
Equation~\eqref{eq:conc_near_wall_explicit} demonstrates a characteristic feature of the cylindrical diffuse layer: near the highly curved interface, the counterion accumulation becomes independent of the cylinder radius $R$. The spatial profile is governed by the electrostatic exponent $2\xi$ and the bulk Debye length $\lambda_D$.
This power-law behavior confirms that the cylindrical curvature suppresses the non-linear exponential terms of the Poisson--Boltzmann equation, maintaining a continuous, non-singular screening profile from the cylinder surface into the bulk electrolyte.

Equation~\eqref{eq:conc_near_wall_explicit} highlights a key distinction between the present approach and Manning’s salt-free theory~\cite{manning.gs:1969} within the subcritical regime ($\xi < 1$). While the infinitely thin line charge in the Manning model leads to a mathematical divergence of the electrostatic potential and ion concentrations at the axis, the finite radius $R$ in the current continuum formulation acts as a physical cutoff. Consequently, the counterion density at the cylinder surface remains strictly finite. The strong local accumulation of ions near the wall as $\xi$ approaches unity is not a precursor to a condensed phase, but rather a regular power-law localization driven by the cylindrical geometry.

\subsubsection{The Non-Linear Boltzmann Explosion}\label{subsec:boltzmann_explosion}

To determine where the linear Laplace approximation fails, we analyze the magnitude of the screening term. At the cylinder surface ($r=R$), the potential is given by Eq.~\eqref{HC_cylinder}, and for $\phi \gg 1$ the hyperbolic sine approaches the exponential limit $\sinh\phi \simeq e^{\phi}/2$. Under these conditions, the screening term remains a small perturbation in the inner zone if:
\begin{equation}\label{eq:screening_scale_cylinder}
\frac{1}{2} \left(\frac{R}{\lambda_D}\right)^2 \left( \frac{2\lambda_D}{r e^{\gamma}} \right)^{2 \xi}  \ll 1.
\end{equation}
In terms of the small parameter $\varepsilon = R/\lambda_D \ll 1$, this expression scales at the interface as $\mathcal{O}(\varepsilon^{2-2\xi})$.

When the charge density is subcritical ($\xi < 1$), the positive exponent ensures that this term vanishes as $\varepsilon \to 0$, validating the Laplacian approximation. At the critical threshold $\xi = 1$, the exponent vanishes. In the supercritical regime ($\xi > 1$), the exponent becomes negative, causing the screening term to diverge as $\varepsilon \to 0$. This singular behavior marks the onset of the non-linear Boltzmann explosion, where the electrolyte screening can no longer be neglected near the cylinder.

Crucially, the factor $(R/r)^{2\xi}$ shows that this non-linear divergence is strictly localized. The Boltzmann explosion forms a sharp electrostatic zone trapped near the cylinder surface. As the radial distance $r$ increases, the local counter-ion density rapidly loses its singular character, transitioning smoothly into the regular diffuse layer governed by the background salt.

\subsection{The Supercritical Regime}\label{subsec:supercritical}

\subsubsection{Matching Procedure for Strongly Non-Linear Fields}\label{subsec:mae2}

In the supercritical regime ($\xi > 1$), the invariance of the far-field potential emerges not from a physical condensation of ions, but as an intrinsic mathematical property driven by the breakdown of geometric linearization. When the charge strength exceeds the critical threshold, the cylindrical curvature can no longer suppress the exponential Boltzmann growth, and linearization fails in the inner region ($\rho \sim 1$). The governing equation in this non-linear inner zone transitions into:
\begin{equation}
\label{eq:app_PB_inner}
\frac{d^2\phi^{(i)}}{d\rho^2} + \frac{1}{\rho}\frac{d\phi^{(i)}}{d\rho} \simeq \frac{1}{2}\varepsilon^2 e^{\phi^{(i)}}.
\end{equation}
Although the explicit prefactor $\varepsilon^2$ is small, the potential itself scales as $\phi^{(i)} \sim -2\ln\varepsilon$ in this zone. This logarithmic growth balances the $\varepsilon^2$ prefactor, maintaining the screening term at an $\mathcal{O}(1)$ scale and ensuring a self-consistent non-linear inner solution.

In the supercritical regime ($\xi > 1$), the counterion density is localized so close to the interface that the effective thickness of the screening cloud becomes much smaller than the cylinder radius ($r - R \ll R$). In this near-surface limit ($\rho \to 1$), the curvature term becomes subdominant compared to the steep gradient of the second derivative. Consequently, the leading-order inner problem reduces to a planar Liouville equation, yielding the solution:
\begin{equation}
\phi^{(i)}(\rho) \simeq \phi_s - 2 \ln \left[ 1 + \xi (\rho - 1) \right],
\label{eq:app_inner_asymp_fixed}
\end{equation}
where the integration constant $\phi_s$ represents the leading-order surface potential. At the cylinder interface ($\rho = 1$), the logarithmic correction term vanishes, leaving $\phi_s$ to be determined by matching this near-field solution with the far-field expansion.

Far from the axis where $\varepsilon\rho \sim 1$, the electrostatic potential decays sufficiently for the Debye–Hückel linearization to become valid. The outer linear solution satisfying the boundary condition at infinity yields the Macdonald profile. To match this solution with the inner zone, we project $\phi^{(o)}$ into the intermediate domain ($\varepsilon\rho \to 0$), utilizing the small-argument expansion given by Eq.~\eqref{eq:match_out_simplified}.

According to the matching principle, the inner and outer representations must become identical within the intermediate overlap domain. Evaluating the non-linear inner profile from Eq.~\eqref{eq:app_inner_asymp_fixed} in the limit $\xi(\rho-1) \gg 1$ yields the asymptotic field $\phi_s - 2\ln\xi - 2\ln\rho$. Matching the coefficients of the singular $\ln\rho$ terms between this inner limit and the far-field expansion in Eq.~\eqref{eq:match_out_simplified} requires $-2 \ln\rho \equiv -M \ln\rho$, which immediately locks the far-field amplitude to a constant value of $M = 2$. Furthermore, requiring the remaining constant terms in the overlap domain to be equal yields the closed-form relation for the leading-order surface potential:
\begin{equation}
\phi_s \simeq 2\ln\left(\frac{2\xi}{\varepsilon}\right) - 2\gamma.
\label{eq:surface_potential_matched_value}
\end{equation}
Thus, in this regime $\phi_s$ also grows on increasing $\xi$, but now weakly logarithmically.

\subsubsection{Composite Global Solution in the Supercritical Regime}
\label{subsec:supercritical_composite}

A uniformly valid analytical profile across the entire domain is constructed by applying the composition rule~\eqref{eq:prandtl}. By combining the non-linear inner solution and the linear far-field function, and subtracting their common intermediate overlap representation, the spatial transitions are preserved. Collecting the remaining logarithmic scales, the uniform composite electrostatic potential is given by:
\begin{equation}
\phi(\rho)\simeq 2 \left[ K_0(\varepsilon \rho) + \ln \left( \frac{\xi \rho}{1 + \xi (\rho - 1)} \right) \right].
\label{eq:composite_solution_final_compact}
\end{equation}
The first term is associated with the long-range linear Debye–Hückel screening driven by the bulk electrolyte. The logarithmic term captures the non-linear counterion accumulation within the near-field cloud.

Substituting the composite potential \eqref{eq:composite_solution_final_compact} into the Boltzmann distribution yields the normalized counterion concentration:
\begin{equation}
\frac{n_-(\rho)}{n_\infty} \simeq \left[ \frac{\xi \rho}{1 + \xi (\rho - 1)} \right]^2 \exp\left[ 2 K_0\left( \varepsilon \rho \right) \right].
\label{eq:counterion_density_prandtl_final}
\end{equation}
As $\rho \to \infty$, the bracketed prefactor approaches unity and $K_0(\varepsilon\rho) \to 0$, ensuring that the concentration satisfies the bulk electroneutrality condition $\lim_{\rho \to \infty} (n_-/n_\infty) = 1$. In the vicinity of the charged surface where $\rho \to 1$, expanding the modified Bessel functions yields the leading-order near-wall asymptotic behavior for the supercritical regime:
\begin{equation}
\frac{n_-(\rho)}{n_\infty} \simeq \xi^2 \left( \frac{2 e^{-\gamma}}{\varepsilon} \right)^2 \left( \frac{1}{\rho} \right)^{2\xi} = \left( \frac{2\xi e^{-\gamma} \lambda_D}{R} \right)^2 \left( \frac{1}{\rho} \right)^{2\xi}.
\label{eq:counterion_density_precise_asymptotic}
\end{equation}
Comparing this result with the subcritical profile~\eqref{eq:conc_near_wall_explicit} shows that their spatial decay is identical, and the supercritical solution differs only by the multiplier $\xi^2 (2e^{-\gamma}/\varepsilon)^{2-2\xi}$.

This scaling comparison reveals a fundamental physical property of the cylindrical diffuse layer under strong electrostatic coupling, providing a direct continuum analogue to Manning’s microscopic partition function $\int r^{1-2\xi} dr$~\cite{manning.gs:1969}. In the ideal line-charge limit ($R \to 0$), the partition function undergoes a thermodynamic instability at $\xi = 1$ due to the divergence at the lower integration limit. In the present continuum framework, however, the finite cylinder radius $R$ serves as a physical cutoff that prevents an unphysical infinite collapse. At the critical threshold ($\xi = 1$), the scaling of both near-wall density profiles continuously converges, with the divergence of the small parameter strictly locked at $\varepsilon^{-2}$. Beyond this threshold ($\xi > 1$), the non-linear governance of the inner Liouville zone completely freezes the exponent of the geometric divergence at this fixed value of $-2$. Consequently, instead of a non-analytic counterion condensation, any further increase in the surface charge density does not intensify the singular spatial scaling of the screening cloud, but is instead accommodated entirely by the smooth quadratic prefactor $\xi^2$.

\subsection{Effective Charge Locking}
\label{subsec:charge_locking}

To evaluate how the non-linear electrostatic effects manifest in the bulk electrolyte, the composite potential can be cast into a representation featuring a spatially varying effective amplitude paired with the linear screening solution:
\begin{equation}
\phi(\rho) = 2 \mathcal{A}(\rho; \xi, \varepsilon) K_0(\varepsilon \rho).
\label{eq:effective_charge_presentation}
\end{equation}
By matching Eq.~\eqref{eq:effective_charge_presentation} with the uniform descriptions derived above, this effective screening factor can be isolated for each electrostatic regime. In the subcritical regime ($\xi < 1$), the effective amplitude loses its spatial dependence and directly tracks the charge parameter:
\begin{equation}
\mathcal{A} \simeq \xi.
\label{eq:effective_charge_sub}
\end{equation}

Conversely, in the supercritical regime ($\xi > 1$), comparing Eq.~\eqref{eq:effective_charge_presentation} with the uniform solution in Eq.~\eqref{eq:composite_solution_final_compact} isolates the effective screening factor as:
\begin{equation}
\mathcal{A} \simeq 1 + \frac{\ln \left( \frac{\xi \rho}{1 + \xi (\rho - 1)} \right)}{K_0(\varepsilon \rho)}.
\label{eq:effective_charge}
\end{equation}

From a physical standpoint, for a thin cylinder ($\varepsilon \ll 1$), we expect the effective charge to stabilize and form a flat plateau in the intermediate region ($1 \ll \rho \ll 1/\varepsilon$). In this domain, the potential has already become small enough for the system to behave linearly, but the exponential screening has not yet set in. However, Eq.~\eqref{eq:effective_charge} fails to capture this behavior. Instead of showing a plateau, the formula grows monotonically in this region and eventually diverges at large distances. This contradiction between physical intuition and the leading-order result leads to two key conclusions:

First, the structure of Eq.~\eqref{eq:effective_charge} reveals that the leading order matching is inherently insufficient to capture a physical plateau, which requires a vanishing radial gradient, $\partial \mathcal{A}/\partial \rho = 0$. In the wide intermediate zone ($1 \ll \rho \ll 1/\varepsilon$), where $K_0(\varepsilon \rho) \approx \ln(2/\gamma \varepsilon) - \ln\rho$, differentiating Eq.~\eqref{eq:effective_charge} shows that the gradient is dominated by the coordinate dependence of this screening denominator rather than the near-field geometry. This forces the derivative to remain strictly non-zero, meaning the leading order solution grows monotonically without stabilizing. Capturing a real plateau requires higher-order corrections that incorporate the near-field potential curvature into the denominator to balance out this screening gradient.

Second, despite its divergence, Eq.~\eqref{eq:effective_charge} provides a useful qualitative estimate for the effective charge in the intermediate region before the divergence sets in. By evaluating Eq.~\eqref{eq:effective_charge} at a characteristic scale $\rho \sim 1/\varepsilon$, where the numerator approaches $\ln[\xi/(\xi-1)]$, we can see that the amplitude of the correction is strictly distinct from unity and is a monotonically increasing function of $\varepsilon$ (since a smaller $\varepsilon$ increases the logarithmic denominator). However, because the numerator is strictly positive, Eq.~\eqref{eq:effective_charge} always predicts a value greater than 1. It fails to capture the regime where the effective charge drops below unity at small $\varepsilon$, highlighting another critical limitation: higher-order corrections must not only stabilize the divergence but also renormalize the numerator to allow values below 1.

Ultimately, these limitations indicate that the amplitude $\mathcal{A}$, while serving as a standard convention within traditional polymer physics, defines the effective charge strictly from a surface-centric perspective. From the standpoint of the bulk electrolyte and double-layer theory, the true non-linear screening response is instead governed by the invariant ratio of the fluid screening scale to the charge density scale, $\lambda_D/\ell_{GC} = \xi/\varepsilon$. Shifting the physical interpretation toward this ratio clarifies that the far-field effective charge is an emergent property of the electrolyte environment, rather than a fixed parameter bound to the geometric boundary of the polymer backbone.

Given that higher-order analytical corrections face the non-uniform convergence typical of non-linear cylindrical frameworks—often correcting the near-field at the expense of far-field physicality—we restrict our analytical treatment to the leading-order matching. This limitation further underscores the necessity of full non-linear numerical calculations, which bypass these asymptotic constraints to establish the existence and precise scaling of the physical plateaus.

\section{Analytical vs. Numerical Results: Unified Continuum Profiles}\label{sec:results}

To validate the analytical derivation, the leading-order composite solutions are compared with numerical profiles of the full nonlinear cylindrical Poisson--Boltzmann equation in an truncated open domain. The numerical benchmarks are obtained using a fourth-order collocation method with adaptive mesh refinement~\cite{bader.g:1987}. To resolve the steep potential gradient near the wall, the equation is formulated in the logarithmic coordinate $s = \ln(r/R)$, which provides an intrinsically dense grid at the surface. The far-field condition is imposed at $r_{\max} = 40\,\lambda_D$ via an asymptotic Robin boundary condition derived from the $K_0$ tail of the linearized solution. The resulting nonlinear algebraic system is solved using a damped Newton iteration. Finally, the far-field amplitude $\mathcal{A}_\infty(\varepsilon)$ is stable to within $10^{-6}$ across independent variations of the grid and solver tolerance, with two independent extraction formulas---based on the potential and its radial derivative---agreeing to within the same accuracy.

\subsection{Potentials and Concentrations}\label{subsec:potentials}

\begin{figure}[h]
\centering
\includegraphics[width=\columnwidth]{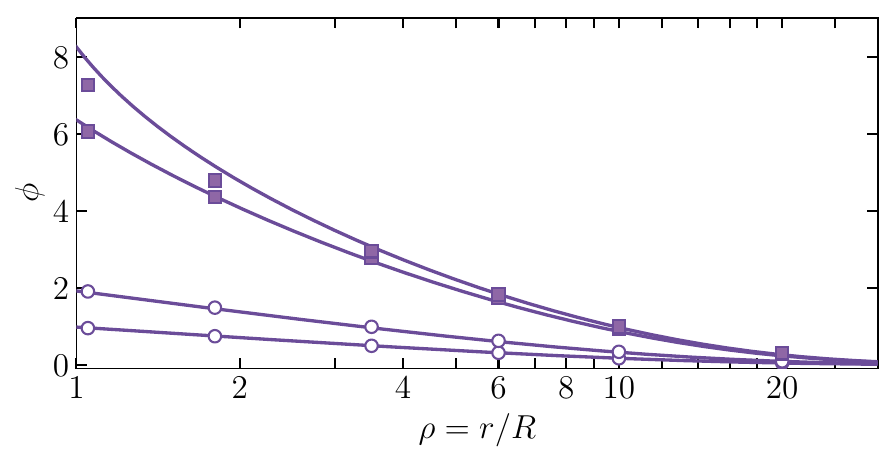}
\caption{Electrostatic potential profiles $\phi$ near a charged cylinder plotted as a function of the dimensionless radial distance $\rho = r/R$ for varying linear charge parameters ($\xi = R/\ell_{GC} = 4, 2, 0.4, 0.2$, from top to bottom) at a fixed screening parameter $\varepsilon = R/\lambda_D = 0.1$. Symbols show calculations from the composite analytical solutions~\eqref{eq:composite_final} and \eqref{eq:composite_solution_final_compact}, while solid lines represent the exact numerical collocation results.}
\label{fig:unified_profiles}
\end{figure}

Figure~\ref{fig:unified_profiles} illustrates the continuous decay of the dimensionless electrostatic potential $\phi$ as a function of the dimensionless radial distance $\rho = r/R$ at $\varepsilon = 0.1$, spanning both the subcritical and supercritical regimes. For subcritical charges ($\xi = 0.2$ and $\xi = 0.4$), the analytical representations are governed by the linear Debye–H\"uckel limit. The calculations from Eq.~\eqref{eq:composite_final} overlay the numerical curves, confirming that cylindrical curvature weakens the non-linear coupling at moderate surface charge densities.
The accuracy of the matched asymptotic derivation is shown upon crossing into the supercritical regime ($\xi = 2.0$ and $\xi = 4.0$), where the linear solution is invalid near the cylinder wall due to the non-linear Boltzmann accumulation. Here, the supercritical composite solution~\eqref{eq:composite_solution_final_compact} captures the smooth crossover toward the Debye–H\"uckel tail. At the cylinder surface ($\rho = 1$), the analytical potential reduces to the value given by Eq.~\eqref{eq:surface_potential_matched_value}. For $\xi = 2.0$, the analytical profile coincides with the numerical results. At the highly supercritical limit of $\xi = 4.0$, where counterion compression produces a steep localized gradient, the leading-order analytical profile exhibits a minor underestimation of approximately $7.5\%$ at the surface. This localized variance is a consequence of omitting the curvature term from the non-linear Poisson--Boltzmann equation in the near-field zone. While neglecting this curvature source term enables a closed-form flat-wall analytical solution, the subsequent asymptotic matching with the outer cylindrical Bessel function $K_0(r/\lambda_D)$ involves the Euler–Mascheroni constant $\gamma$, leading to this minor numerical underprediction at the wall.

\begin{figure}[h]
\centering
\includegraphics[width=\columnwidth]{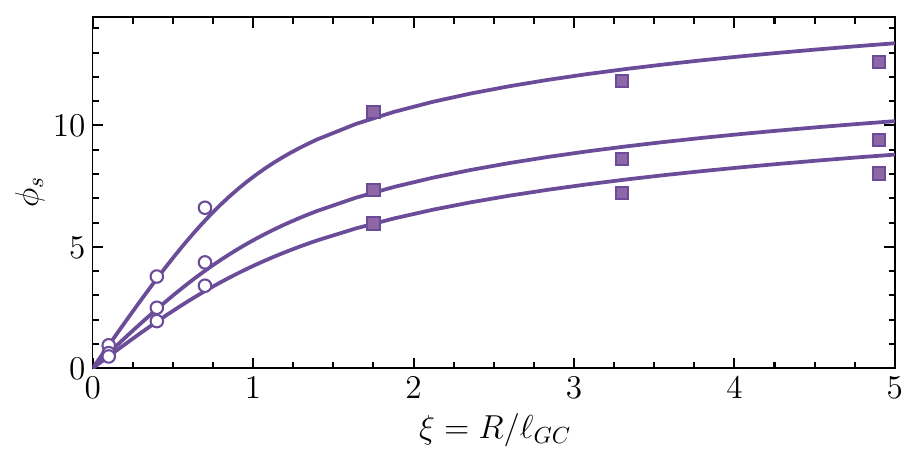}
\caption{Surface potential $\phi_s$ as a function of the dimensionless charge parameter $\xi$ computed for fixed $\varepsilon = 0.01, 0.05, 0.1$ (from top to bottom). Symbols are calculated from Eqs.~\eqref{HC_cylinder} and \eqref{eq:surface_potential_matched_value}, while solid lines represent the exact numerical collocation results.}
\label{fig:boltzmann_explosion}
\end{figure}

These profiles refer to fixed $\xi$ and varying $\rho$. Fixing $\rho = 1$ for several $\varepsilon \ll 1$ while varying $\xi$ corresponds to the curves in Fig.~\ref{fig:boltzmann_explosion}. The plot tracks the breakdown of the subcritical approximation~\eqref{HC_cylinder} as $\xi \to 1$. The exact numerical curves grow monotonically across the range of $\xi$, demonstrating that the crossover from the subcritical to the supercritical regime is continuous, with no jumps or phase transitions.
The variation of $\xi$ is plotted up to $\xi = 5$. This range is chosen to encompass all practically relevant polyelectrolytes, as the charge parameter of real macromolecules—including highly charged systems such as double-stranded DNA ($\xi \simeq 4.2$)—is bounded below this limit. Extending the axis to $\xi = 5$ thus captures the full physical transition zone while clearly revealing the convergence behavior toward the supercritical asymptotic plateaus.
The analytical linear Eq.~\eqref{HC_cylinder} coincides with the exact non-linear Poisson--Boltzmann solution below $\xi \simeq 0.75$, even when the surface potential reaches $\phi_s \simeq 7$. Above this threshold, the linear equation overestimates the potential, indicating that as the system approaches the critical region, the non-linear accumulation of counterions near the wall provides enhanced electrostatic screening that suppresses further potential growth.
Conversely, the supercritical Eq.~\eqref{eq:surface_potential_matched_value} follows the numerical curve above $\xi \simeq 1.5$, yielding only a small discrepancy, with a slight overestimation below $\xi \simeq 2$ and a slight underestimation above it. This agreement is obtained using a leading-order derivation only. Without invoking higher-order curvature corrections, this compact leading-order asymptotic expression captures the quantitative behavior of the surface potential over the physical range of supercritical charge parameters.

\begin{figure}[h]
\centering
\includegraphics[width=\columnwidth]{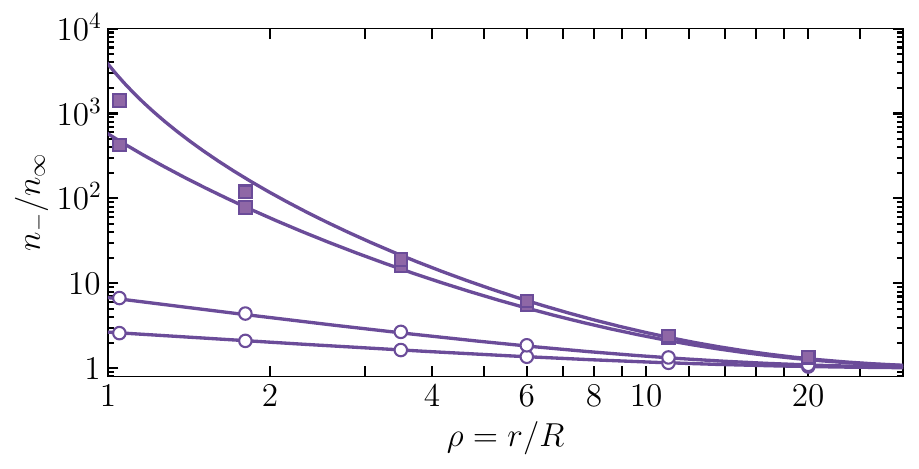}
\caption{Continuous counterion concentration profiles $n_-(r)/n_\infty$ surrounding the cylinder calculated using $R/\lambda_D = 0.1$ for $\xi = 4, 2, 0.4, 0.2$ (from top to bottom). Open and filled symbols show calculations from Eqs.~\eqref{eq:conc_near_wall} and \eqref{eq:counterion_density_prandtl_final}, respectively, while solid lines represent the exact numerical collocation results.}
\label{fig:concentration_profiles}
\end{figure}

The physical consequences of these electrostatic profiles are reflected in the distribution of mobile ions. Figure~\ref{fig:concentration_profiles} displays the continuous counterion concentration profiles $n_-(r)/n_\infty$ as a function of the dimensionless radial distance $\rho = r/R$ on a semi-logarithmic scale. For subcritical parameters ($\xi = 0.2$ and $\xi = 0.4$), the analytical representations governed by Eq.~\eqref{eq:conc_near_wall} overlay the numerical curves. Upon entering the supercritical regime ($\xi = 2.0$ and $\xi = 4.0$), the exponential Boltzmann coupling leads to a large accumulation of counterions near the cylinder surface. The smoothly varying nature of these profiles across all values of $\xi$ confirms that the crossover between the regimes occurs continuously. In contrast to the step-function approximations of the classical Manning threshold model, the full Poisson--Boltzmann solution for a finite-radius cylinder resolves a smooth, non-singular electrostatic profile without unphysical jump discontinuities.
Despite the steep concentration profiles spanning several orders of magnitude, the leading-order analytical representations~\eqref{eq:counterion_density_prandtl_final} maintain quantitative accuracy, capturing both the near-wall accumulation and the decay toward the bulk limit.

\subsection{The Myth of Effective Charge Saturation}\label{subsec:charges}

\begin{figure}[h]
	\centering
	\includegraphics[width=\columnwidth]{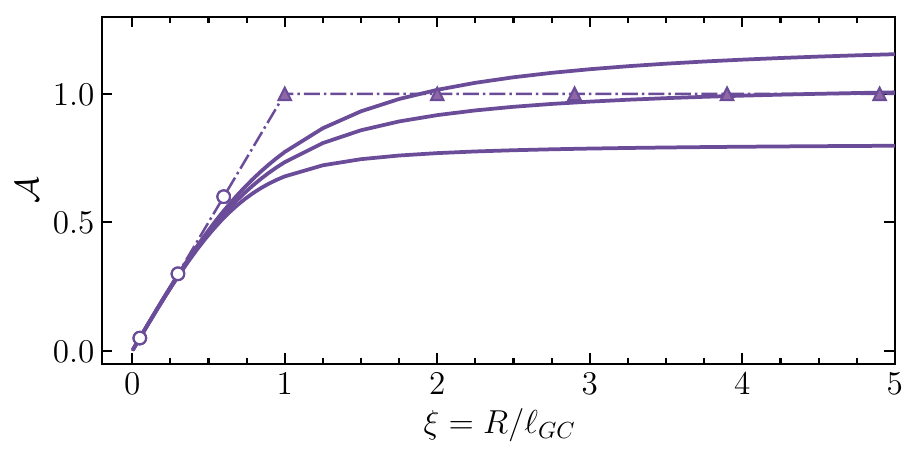}
	\caption{Effective charge $\mathcal{A}(\xi)$ inferred from the numerical solution of the full non-linear Poisson--Boltzmann equation, plotted for fixed screening parameters $\varepsilon = R/\lambda_D = 0.1, 0.05, 0.01$ (from top to bottom). Open circles show the subcritical Debye–H\"uckel prediction calculated from Eq.~\eqref{eq:effective_charge_sub}. Dash-dotted lines represent the Manning threshold values, with $\mathcal{A}=1$ marked by filled triangles.}
	\label{fig:amplitude_locking}
\end{figure}

The behavior of the far-field amplitude $\mathcal{A}(\xi)$ across varying electrolyte concentrations ($\varepsilon = 0.1, 0.05, 0.01$) is shown in Fig.~\ref{fig:amplitude_locking}. When plotted against the charge parameter $\xi$, the non-linear curves converge to the subcritical Debye–H\"uckel prediction (open circles) at small $\xi$. This agreement confirms that the system follows the linear Bessel-function solution described by Eq.~\eqref{eq:effective_charge_sub}, even when the electrostatic potential at the cylinder surface is large. The non-linear numerical solutions track this linear trend up to $\xi \simeq 0.75$, rather than extending to the classical Manning threshold $\xi = 1$.
Beyond $\xi \simeq 0.75$, the non-linear Boltzmann factor alters the electrostatic response, shifting the system into a precritical regime where the curves smoothly break away from the linear prediction. Instead of undergoing a sharp, non-analytic transition at $\xi = 1$ or locking onto a universal threshold $\mathcal{A} = 1$, the effective charge levels off continuously. Upon entering the supercritical regime ($\xi > 1$), the amplitude does not jump to a constant value but gradually approaches distinct, concentration-dependent parallel plateaus. This transition zone reflects a continuous spatial crossover within the diffuse layer, rather than a sudden on-surface pinning event.

The screening parameter $\varepsilon$ determines the rate of this non-linear crossover within the range of physical charge parameters ($\xi \le 5$). For low salt content ($\varepsilon = 0.01$), the curve flattens rapidly, establishing a parallel plateau early in the supercritical regime. At moderate salinity ($\varepsilon = 0.05$), the profile approaches its asymptotic level by $\xi \simeq 5$. Conversely, for higher salt concentrations ($\varepsilon = 0.1$), the system remains far from saturation within this physical window; while the plateau mathematically exists, it is shifted toward unphysically large values of $\xi$. A systematic decrease in the salt content ($\varepsilon \to 0.01$) induces a monotonic downward shift of these supercritical plateaus, forcing the accessible effective charge to stabilize below the classical Manning limit.

\begin{figure}[h]
\centering
\includegraphics[width=\columnwidth]{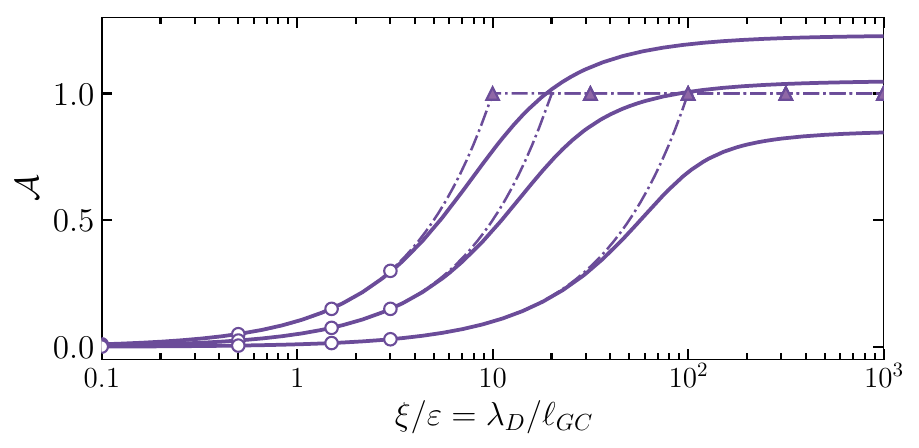}
\caption{Far-field effective charge $\mathcal{A}$ mapped onto the scale $\lambda_D/\ell_{GC} = \xi/\varepsilon$, computed for fixed dimensionless cylinder radii $\varepsilon = 0.1, 0.05, 0.01$ (from top to bottom). Open circles show the subcritical prediction governed by Eq.~\eqref{eq:effective_charge_sub}, while filled triangles mark the classical Manning threshold ($\mathcal{A}=1$). The curves emerge from the origin with distinct slopes proportional to $\varepsilon$, pass through smooth inflection points, and stabilize into separate, parallel plateaus that preserve a permanent logarithmic memory of the cylinder curvature.}
\label{fig:A_eff_graham_scaling}
\end{figure}

To test whether the saturation of the effective charge is governed by a universal scale or remains coupled to the macroion geometry, the amplitude $\mathcal{A}$ is mapped onto the parameter $\lambda_D/\ell_{GC} = \xi/\varepsilon$. In colloid and interface science, this ratio serves as a metric for the onset of non-linear screening. Tracking $\mathcal{A}$ across this parameter (Fig.~\ref{fig:A_eff_graham_scaling}) shows whether the asymptotic saturation levels collapse into a universal scaling curve, or if the cylindrical geometry enforces distinct behaviors.
The resulting curves in Fig.~\ref{fig:A_eff_graham_scaling} contrast with the behavior of the surface potential $\phi_s$. At strong charging ($\xi/\varepsilon \gg 1$), the counterion cloud compresses into a microscopic layer immediately adjacent to the cylinder wall. For these counterions, the local potential $\phi_s$ effectively ``forgets'' the curvature, causing the surface potential curves to approach the flat-wall limit.

Metaphorically, this behavior resembles the transition from a ground-level observer standing next to a cylinder—where the immediate surface appears flat—to a vantage point from the far field that resolves the global cylindrical geometry. Unlike the near-field potential, the effective charge $\mathcal{A}$ is determined in the outer matching zone, where the background electrolyte screens the macroion. Because the long-range electrostatic lines of force remain cylindrical, the far-field asymptotic tail governed by the Macdonald function $K_0(\varepsilon \rho)$ preserves a permanent memory of the cylinder curvature $\varepsilon$. Instead of collapsing onto a single universal curve, the effective charge splits into a series of distinct, parallel plateaus. While near-field potentials can be locally regularized at high densities, the true far-field effective charge of a cylinder emerges as an inherent property that cannot be decoupled from its geometry.

This continuous spatial crossover is characterized by the distinct inflection points visible on the S-shaped curves in Fig.~\ref{fig:A_eff_graham_scaling}. Physically, each inflection point marks the position where the non-linear Boltzmann accumulation---identified via the asymptotic scaling $\mathcal{O}(\varepsilon^{2-2\xi})$ in Eq.~\eqref{eq:screening_scale_cylinder}---reaches its maximum rate of spatial growth. The horizontal displacement of these inflection points mirrors the structure of the screening threshold. For moderate salinity ($\varepsilon = 0.1$), the pre-exponential factor in Eq.~\eqref{eq:screening_scale_cylinder} is large, causing the non-linear electrolyte response to activate early on the scaling axis. Conversely, for thin cylinders in the dilute regime ($\varepsilon = 0.01$), the small cylinder curvature suppresses the onset of screening, forcing the inflection point to shift to the right. In this case, the system must reach a much higher value of the parameter $\xi/\varepsilon$ before the non-linear near-field zone can fully develop and drive the curve toward its parallel plateau.

Because the parameter $\varepsilon$ is kept fixed for each individual curve, extending the scaling axis toward large values ($\xi/\varepsilon \simeq 1000$) forces the charge parameter $\xi$ to grow beyond physical limits. In experiments, the charge parameter of a macromolecular backbone is bounded, typically within $\xi \le 5$~\cite{manning.gs:1969}. Consequently, for a highly screened system ($\varepsilon = 0.1$), the physically accessible domain terminates at $\xi/\varepsilon = 50$, well before the onset of the asymptotic plateau. Conversely, for the low-salinity regime ($\varepsilon = 0.01$), the realistic window extends up to $\xi/\varepsilon = 500$, meaning that the effective charge stabilizes into its parallel plateau within attainable experimental conditions.

It is historically instructive to re-examine the original formulation by Manning~\cite{manning.gs:1969}, which subsequent computational literature has rigidified into an invariant view of absolute charge locking ($\mathcal{A} = 1$). In his seminal paper, Manning explicitly recognized the highly idealized nature of the salt-free, infinitely thin line-charge limit, openly stating that the residual effective charge in real electrolyte environments could float either above or below unity~\cite{manning.gs:1969}. The two-phase step-function threshold was introduced primarily as a heuristic mathematical convenience to regularize the limiting laws of polyelectrolyte solutions. The numerical and analytical results in Fig.~\ref{fig:amplitude_locking} and Fig.~\ref{fig:A_eff_graham_scaling} resolve these foundational hesitations. By accounting for the regularizing effect of the background salt within a continuous mean-field near-field description, the effective charge is shown to be salt-dependent. The separate, parallel supercritical plateaus do not violate Manning's original intuition; rather, they provide the exact, parameter-free mathematical realization of the physical boundaries he foresaw over half a century ago.

\section{Fallacy of Conventional Paradigms}
\label{sec:paradigms}

\subsection{Counterion Condensation as a Spatial Crossover}
\label{subsec:condensation_fallacy}

The composite analytical solution and numerical verification allow us to analyze the conceptual limitations of conventional polyelectrolyte electrostatics. For decades, literature influenced by historical salt-free approximations---ranging from cylindrical cell models to line-charge condensation approaches---has treated the cylindrical geometry as an exceptional case requiring artificial mechanisms such as step-like counterion condensation or heuristic effective charges~\cite{deserno.m:2000}. Our derivation demonstrates that this behavior is uniform across all charge regimes. In the presence of a background electrolyte, the charged cylinder loses its mathematical isolation. In the subcritical regime ($\xi < 1$), the potential is governed globally by the unmodified Bessel function $K_0(\varepsilon \rho)$ using the true surface charge. Near the surface ($\rho = 1$), the cylindrical curvature suppresses the non-linear coupling, forcing the counterions to remain diffuse and continuously distributed according to a smooth power-law scaling. There is no phase-like localization or step-like collapse of ions onto the cylinder surface, rendering the concept of a renormalized effective charge unnecessary for moderate fields.

The breakdown of this linear model at $\xi = 1$ is not a sudden, non-analytic phase transition as suggested by subsequent computational literature. Manning's original formulation identified $\xi = 1$ as the threshold where electrostatic energy balances thermal fluctuations ($k_B T$), providing a vital physical insight for polymer physics~\cite{manning.gs:1969}. However, Manning explicitly recognized the highly idealized nature of the salt-free, infinitely thin line-charge limit, openly noting that the residual effective charge in real electrolyte environments could float either above or below unity~\cite{manning.gs:1969}. Our analytical solution resolves these foundational hesitations. The breakdown represents the physical onset of strong non-linear near-field effects where the exponential Boltzmann scaling overwhelms geometric dilution. When the system crosses into the supercritical regime ($\xi > 1$), the far-field effective charge $\mathcal{A}$ does not abruptly freeze at a universal value. Instead, the charge plateau emerges as a direct mathematical consequence of the integration between the near-field flat-wall limit and the outer Debye screening tail. Because this plateau is regularized by the background electrolyte, its height is sensitive to the screening parameter $\varepsilon$, allowing it to shift continuously and drop below the unperturbed level---a feature that eludes rigid threshold models.

This spatial redistribution clarifies the limits of historical approximations. For rough equilibrium estimates in macroscopic chemistry---such as calculating osmotic pressure or global activity coefficients---Manning's heuristic shortcut functions as a reasonable thermodynamic proxy, as it captures the integrated counterion density within the diffuse layer~\cite{manning.gs:1969}. However, when applied to dynamic transport phenomena and electrokinetics, this approach fails. In non-equilibrium electrophoresis, the driving force propelling the charged cylinder is generated by hydrodynamic shear stresses at the wall, driven by the local velocity gradients within the diffuse layer. By removing the near-field counterions from the continuous non-linear transport equations and declaring them an immobile, bound mass, the traditional view distorts the volume force distribution near the interface. Recent advancements in nanofluidics support this continuous picture, demonstrating that the intrinsic mobility of near-wall ions actively amplifies electrokinetic flows and enhances the effective zeta-potential~\cite{asmolov.es:2025b}. Stripping the hydrodynamic equations of this driving mechanism renders the rigid, step-like condensation approach non-applicable to electrokinetic transport.

Consequently, multi-parameter numerical models that introduce empirical effective charges to fit experimental data become unnecessary. The background salt regularizes the system from within, bridging the physics of the cylindrical double layer with that of planar boundaries. Throughout all regimes, the counterion atmosphere remains diffuse, mobile, and mathematically continuous, proving that the limiting laws of counterion condensation are artifacts of omitting the screening electrolyte and neglecting the non-linear near-field structure.

Finally, these findings can be grounded in physical systems. In experiments, the Manning parameter rarely exceeds $\xi \simeq 4$, a structural ceiling dictated by the steric limits of polymer backbones and the Bjerrum length of water ($\ell_B \simeq 0.71$~nm). For instance, double-stranded DNA represents one of the most highly charged natural polyelectrolytes, yet its structural charge density is capped at $\xi \simeq 4.2$. Our analytical solution demonstrates that the strict far-field charge locking postulated by salt-free models does not apply here. The horizontal plateau of the effective charge is a mathematical asymptote that only solidifies when $\xi \ge 5$. Because physical macromolecules are constrained to the $\xi \le 4$ regime, they never reach this idealized plateau. Instead, they operate entirely within a smooth, continuous transition zone where the effective charge remains a dynamic function of both the chemical charge density and the background salinity. This behavior renders the rigid, step-like models of counterion condensation non-applicable to real electrolyte solutions.

\subsection{Perspectives on Electrostatic Stiffness and Persistence Length}\label{subsec:stiffness}

The finding that the far-field effective charge $\mathcal{A}(\xi, \varepsilon)$ does not lock onto a rigid, universal threshold but instead stabilizes at distinct, salt-dependent parallel plateaus has direct implications for the conformational mechanics of charged polymers. This salt dependence offers a way to clarify the forty-year-old controversy surrounding the scaling of the electrostatic persistence length, historically expressed as $l_e \propto \lambda_D^\alpha$. For decades, the literature has remained divided between the quadratic scaling ($\alpha = 2$) of the Odijk--Skolnick--Fixman (OSF)~\cite{odijk.t:1977,skolnick.pe:1977} and Khokhlov--Khachaturian (KK)~\cite{khokhlov.pe:1982} theories, and the linear scaling ($\alpha = 1$) of the Barrat--Joanny (BJ) description~\cite{barrat.pe:1993}. Concurrently, molecular simulations routinely produce a broad scatter of intermediate, shifting exponents that elude conventional theoretical limits.

Our analytical solution shows that this discrepancy arises from treating the effective charge as a constant independent of salt. Because the amplitude $\mathcal{A}(\xi, \varepsilon)$ varies continuously with both the surface charge density $\xi$ and the screening parameter $\varepsilon = R/\lambda_D$, it modifies the net scaling of the bending stiffness. The behavior of the plateau values shifts in opposite directions depending on the salt concentration. At moderately high ionic strengths ($\varepsilon = 0.1$), the effective amplitude surpasses the Manning threshold ($\mathcal{A} > 1.2$), meaning that the classical OSF/KK integrals underpredict the electrostatic repulsion and the resulting chain stiffness. Conversely, in highly dilute, weak-screening regimes ($\varepsilon = 0.01$), the charge plateau drops significantly below unity ($\mathcal{A} < 0.85$). This decrease provides a continuum justification for the suppressed stiffness observed in asymptotic limits---a phenomenon that led to severe artifacts in early Yukawa-based bead-rod simulations~\cite{stevens.mj:1995}.

While large-scale simulations by \citet{everaers.r:2002} recovered the KK quadratic scaling by expanding the simulation box and eliminating finite-length artifacts, they still relied on linear screening approximations. Inserting our salt-dependent effective charge into the classical KK/OSF integrals bridges the gap between competing theories, complementing the scaling diagrams recently proposed by \citet{rumyantsev.epl1:2026} and \citet{gavrilov.epl2:2026}. 
Although the detailed algebraic derivation of the modified bending integrals is deferred to a future study, the continuum scaling link established in this work clarifies the physical mechanism behind the bending stiffness and flexibility of charged macromolecules.

\section{Conclusions}\label{sec:conclusions}

This work was motivated by an awareness of two outstanding mathematical puzzles related to cylindrical geometry. The first of these was the analytical solution missing in the literature for an open domain in a counterion-only case, a gap that led to the formulation and subsequent extensive expansion of so-called cell models. Here, we have derived such a closed-form analytical solution and demonstrated that it unambiguously indicates the physical breakdown of the salt-free model itself. The second problem had to do with the absence of closed-form solutions to the non-linear cylindrical Poisson---Boltzmann equation with added salt. We have constructed a smooth composite analytical solution for a thin cylinder by applying the method of matched asymptotics. Our derivation shows that the ionic atmosphere remains continuous and diffuse across all charge regimes, which renders the counterion condensation hypothesis redundant. Consequently, this approach eliminates the need for both the charge-truncation postulates of the Manning model and the artificial spatial cutoffs imposed by cell models.

Some further comments should be made. Our analysis gives precisely the Manning critical charge parameter, $\xi = 1$, but this does not correspond to ion condensation. This value indicates a transition between the linear subcritical ($\xi < 1$) and nonlinear supercritical ($\xi > 1$) regimes, and this represents a smooth crossover. We emphasize that the subcritical regime is not restricted to low potentials since the high curvature efficiently deactivates the mean-field nonlinearity.
We have also shown that although in the supercritical regime the effective charge freezes on a constant plateau, the height of this plateau is not fixed to unity as commonly expected, but varies with salinity. Furthermore, within the realistic regime where $\xi \le 5$, these asymptotic saturation levels are not always reached.

These results offer a new perspective on the long-standing debate surrounding the electrostatic persistence length of polyelectrolytes. Incorporating this continuous, salt-dependent effective charge into models provides a clear framework for understanding the variations in scaling exponents observed in the literature. By shifting the description of the effective charge from a rigid surface constant to an environment-dependent property, this work opens up routes to describe macromolecular mechanics without relying on empirical fitting parameters. Our findings also provide an analytical foundation for the electrophoresis and interactions of highly charged cylindrical architectures at finite salt concentrations.

\begin{acknowledgments}
 We thank E.~S.~Asmolov and A.~R.~Khokhlov for the critical reading of the manuscript and helpful comments. This work was supported by the Ministry of Science and Higher Education of the Russian Federation.
\end{acknowledgments}

\section*{DATA AVAILABILITY}

The data that support the findings of this study are available within the
article.

\section*{AUTHOR DECLARATIONS}

The authors have no conflicts to disclose.


\begin{thebibliography}{35}%
	\makeatletter
	\providecommand \@ifxundefined [1]{%
		\@ifx{#1\undefined}
	}%
	\providecommand \@ifnum [1]{%
		\ifnum #1\expandafter \@firstoftwo
		\else \expandafter \@secondoftwo
		\fi
	}%
	\providecommand \@ifx [1]{%
		\ifx #1\expandafter \@firstoftwo
		\else \expandafter \@secondoftwo
		\fi
	}%
	\providecommand \natexlab [1]{#1}%
	\providecommand \enquote  [1]{``#1''}%
	\providecommand \bibnamefont  [1]{#1}%
	\providecommand \bibfnamefont [1]{#1}%
	\providecommand \citenamefont [1]{#1}%
	\providecommand \href@noop [0]{\@secondoftwo}%
	\providecommand \href [0]{\begingroup \@sanitize@url \@href}%
	\providecommand \@href[1]{\@@startlink{#1}\@@href}%
	\providecommand \@@href[1]{\endgroup#1\@@endlink}%
	\providecommand \@sanitize@url [0]{\catcode `\\12\catcode `\$12\catcode
		`\&12\catcode `\#12\catcode `\^12\catcode `\_12\catcode `\%12\relax}%
	\providecommand \@@startlink[1]{}%
	\providecommand \@@endlink[0]{}%
	\providecommand \url  [0]{\begingroup\@sanitize@url \@url }%
	\providecommand \@url [1]{\endgroup\@href {#1}{\urlprefix }}%
	\providecommand \urlprefix  [0]{URL }%
	\providecommand \Eprint [0]{\href }%
	\providecommand \doibase [0]{http://dx.doi.org/}%
	\providecommand \selectlanguage [0]{\@gobble}%
	\providecommand \bibinfo  [0]{\@secondoftwo}%
	\providecommand \bibfield  [0]{\@secondoftwo}%
	\providecommand \translation [1]{[#1]}%
	\providecommand \BibitemOpen [0]{}%
	\providecommand \bibitemStop [0]{}%
	\providecommand \bibitemNoStop [0]{.\EOS\space}%
	\providecommand \EOS [0]{\spacefactor3000\relax}%
	\providecommand \BibitemShut  [1]{\csname bibitem#1\endcsname}%
	\let\auto@bib@innerbib\@empty
	\bibitem [{\citenamefont {Fuoss}\ \emph {et~al.}(1951)\citenamefont {Fuoss},
		\citenamefont {Katchalsky},\ and\ \citenamefont {Lifson}}]{fuoss.rm:1951}%
	\BibitemOpen
	\bibfield  {author} {\bibinfo {author} {\bibfnamefont {R.~M.}\ \bibnamefont
			{Fuoss}}, \bibinfo {author} {\bibfnamefont {A.}~\bibnamefont {Katchalsky}}, \
		and\ \bibinfo {author} {\bibfnamefont {S.}~\bibnamefont {Lifson}},\
	}\bibfield  {title} {\enquote {\bibinfo {title} {The potential of an infinite
				rod-like molecule and the distribution of the counter ions},}\ }\href@noop {}
	{\bibfield  {journal} {\bibinfo  {journal} {Proc. Nat. Acad. Sci.}\ }\textbf
		{\bibinfo {volume} {37}},\ \bibinfo {pages} {579} (\bibinfo {year}
		{1951})}\BibitemShut {NoStop}%
	\bibitem [{\citenamefont {Alfrey~Jr}\ \emph {et~al.}(1951)\citenamefont
		{Alfrey~Jr}, \citenamefont {Berg},\ and\ \citenamefont
		{Morawetz}}]{alfrey.t:1951}%
	\BibitemOpen
	\bibfield  {author} {\bibinfo {author} {\bibfnamefont {T.}~\bibnamefont
			{Alfrey~Jr}}, \bibinfo {author} {\bibfnamefont {P.~W.}\ \bibnamefont {Berg}},
		\ and\ \bibinfo {author} {\bibfnamefont {H.}~\bibnamefont {Morawetz}},\
	}\bibfield  {title} {\enquote {\bibinfo {title} {The counterion distribution
				in solutions of rod-shaped polyelectrolytes},}\ }\href@noop {} {\bibfield
		{journal} {\bibinfo  {journal} {J. Polymer Sci.}\ }\textbf {\bibinfo {volume}
			{7}},\ \bibinfo {pages} {543} (\bibinfo {year} {1951})}\BibitemShut {NoStop}%
	\bibitem [{\citenamefont {Le~Bret}\ and\ \citenamefont
		{Zimm}(1984)}]{lebret1984distribution}%
	\BibitemOpen
	\bibfield  {author} {\bibinfo {author} {\bibfnamefont {M.}~\bibnamefont
			{Le~Bret}}\ and\ \bibinfo {author} {\bibfnamefont {B.~H.}\ \bibnamefont
			{Zimm}},\ }\bibfield  {title} {\enquote {\bibinfo {title} {Distribution of
				counterions around a cylindrical polyelectrolyte and {M}anning's condensation
				theory},}\ }\href {\doibase 10.1002/bip.360230209} {\bibfield  {journal}
		{\bibinfo  {journal} {Biopolymers}\ }\textbf {\bibinfo {volume} {23}},\
		\bibinfo {pages} {287} (\bibinfo {year} {1984})}\BibitemShut {NoStop}%
	\bibitem [{\citenamefont {Andelman}(2006)}]{andelman.d:2006}%
	\BibitemOpen
	\bibfield  {author} {\bibinfo {author} {\bibfnamefont {D.}~\bibnamefont
			{Andelman}},\ }\bibfield  {title} {\enquote {\bibinfo {title} {Introduction
				to electrostatics in soft and biological matter},}\ }in\ \href@noop {} {\emph
		{\bibinfo {booktitle} {Soft Condensed Matter Physics in Molecular and Cell
				Biology}}},\ Vol.~\bibinfo {volume} {6}\ (\bibinfo  {publisher} {Taylor \&
		Francis},\ \bibinfo {address} {New York},\ \bibinfo {year} {2006})\ pp.\
	\bibinfo {pages} {97--122}\BibitemShut {NoStop}%
	\bibitem [{\citenamefont {Manning}(1969)}]{manning.gs:1969}%
	\BibitemOpen
	\bibfield  {author} {\bibinfo {author} {\bibfnamefont {G.~S.}\ \bibnamefont
			{Manning}},\ }\bibfield  {title} {\enquote {\bibinfo {title} {Limiting laws
				and counterion condensation in polyelectrolyte solutions {I}. {C}olligative
				properties},}\ }\href@noop {} {\bibfield  {journal} {\bibinfo  {journal} {J.
				Chem. Phys.}\ }\textbf {\bibinfo {volume} {51}},\ \bibinfo {pages} {924}
		(\bibinfo {year} {1969})}\BibitemShut {NoStop}%
	\bibitem [{\citenamefont {Barrat}\ and\ \citenamefont
		{Joanny}(1996)}]{barrat.jl:1996}%
	\BibitemOpen
	\bibfield  {author} {\bibinfo {author} {\bibfnamefont {J.~L.}\ \bibnamefont
			{Barrat}}\ and\ \bibinfo {author} {\bibfnamefont {J.~F.}\ \bibnamefont
			{Joanny}},\ }\bibfield  {title} {\enquote {\bibinfo {title} {Theory of
				polyelectrolyte solutions},}\ }\href@noop {} {\bibfield  {journal} {\bibinfo
			{journal} {Adv. Chem. Phys.}\ }\textbf {\bibinfo {volume} {94}},\ \bibinfo
		{pages} {1} (\bibinfo {year} {1996})}\BibitemShut {NoStop}%
	\bibitem [{\citenamefont {Netz}\ and\ \citenamefont
		{Andelman}(2001)}]{netz.review:2001}%
	\BibitemOpen
	\bibfield  {author} {\bibinfo {author} {\bibfnamefont {R.~R.}\ \bibnamefont
			{Netz}}\ and\ \bibinfo {author} {\bibfnamefont {D.}~\bibnamefont
			{Andelman}},\ }\bibfield  {title} {\enquote {\bibinfo {title}
			{Polyelectrolytes in solution and at surfaces},}\ }in\ \href@noop {} {\emph
		{\bibinfo {booktitle} {Physics of Biomolecules and Cells}}},\ \bibinfo
	{series} {Les Houches Summer School Proceedings}, Vol.~\bibinfo {volume}
	{75},\ \bibinfo {editor} {edited by\ \bibinfo {editor} {\bibfnamefont
			{H.}~\bibnamefont {Flyvbjerg}}, \bibinfo {editor} {\bibfnamefont
			{F.}~\bibnamefont {J{\"u}licher}}, \bibinfo {editor} {\bibfnamefont
			{P.}~\bibnamefont {Ormos}}, \ and\ \bibinfo {editor} {\bibfnamefont
			{F.}~\bibnamefont {David}}}\ (\bibinfo  {publisher} {EDP Sciences /
		Springer-Verlag},\ \bibinfo {address} {Les Ulis / Berlin},\ \bibinfo {year}
	{2001})\ pp.\ \bibinfo {pages} {235--332}\BibitemShut {NoStop}%
	\bibitem [{\citenamefont {Dobrynin}\ and\ \citenamefont
		{Rubinstein}(2005)}]{dobrynin.av:2005}%
	\BibitemOpen
	\bibfield  {author} {\bibinfo {author} {\bibfnamefont {A.~V.}\ \bibnamefont
			{Dobrynin}}\ and\ \bibinfo {author} {\bibfnamefont {M.}~\bibnamefont
			{Rubinstein}},\ }\bibfield  {title} {\enquote {\bibinfo {title} {Theory of
				polyelectrolytes in solutions and at surfaces},}\ }\href@noop {} {\bibfield
		{journal} {\bibinfo  {journal} {Progress in Polymer Sci.}\ }\textbf {\bibinfo
			{volume} {30}} (\bibinfo {year} {2005})}\BibitemShut {NoStop}%
	\bibitem [{\citenamefont {Odijk}(1977)}]{odijk.t:1977}%
	\BibitemOpen
	\bibfield  {author} {\bibinfo {author} {\bibfnamefont {T.}~\bibnamefont
			{Odijk}},\ }\bibfield  {title} {\enquote {\bibinfo {title} {Polyelectrolytes
				near the rod limit},}\ }\href@noop {} {\bibfield  {journal} {\bibinfo
			{journal} {Journal Polymer Sci.}\ }\textbf {\bibinfo {volume} {15}},\
		\bibinfo {pages} {477} (\bibinfo {year} {1977})}\BibitemShut {NoStop}%
	\bibitem [{\citenamefont {Gonz{\'a}lez-Mozuelos}\ and\ \citenamefont {Olvera
			de~la Cruz}(1995)}]{gonzalez.p:1995}%
	\BibitemOpen
	\bibfield  {author} {\bibinfo {author} {\bibfnamefont {P.}~\bibnamefont
			{Gonz{\'a}lez-Mozuelos}}\ and\ \bibinfo {author} {\bibfnamefont
			{M.}~\bibnamefont {Olvera de~la Cruz}},\ }\bibfield  {title} {\enquote
		{\bibinfo {title} {Ion condensation in salt-free dilute polyelectrolyte
				solutions},}\ }\href@noop {} {\bibfield  {journal} {\bibinfo  {journal} {J.
				Chem. Phys.}\ }\textbf {\bibinfo {volume} {103}},\ \bibinfo {pages} {3145}
		(\bibinfo {year} {1995})}\BibitemShut {NoStop}%
	\bibitem [{\citenamefont {Stevens}\ and\ \citenamefont
		{Kremer}(1995)}]{stevens.mj:1995}%
	\BibitemOpen
	\bibfield  {author} {\bibinfo {author} {\bibfnamefont {M.~J.}\ \bibnamefont
			{Stevens}}\ and\ \bibinfo {author} {\bibfnamefont {K.}~\bibnamefont
			{Kremer}},\ }\bibfield  {title} {\enquote {\bibinfo {title} {The nature of
				flexible linear polyelectrolytes in salt free solution: {A} molecular
				dynamics study},}\ }\href@noop {} {\bibfield  {journal} {\bibinfo  {journal}
			{J. Chem. Phys.}\ }\textbf {\bibinfo {volume} {103}},\ \bibinfo {pages}
		{1669} (\bibinfo {year} {1995})}\BibitemShut {NoStop}%
	\bibitem [{\citenamefont {Grosberg}\ and\ \citenamefont
		{Khokhlov}(1994)}]{grosberg.ayu:1994}%
	\BibitemOpen
	\bibfield  {author} {\bibinfo {author} {\bibfnamefont {A.~Y.}\ \bibnamefont
			{Grosberg}}\ and\ \bibinfo {author} {\bibfnamefont {A.~R.}\ \bibnamefont
			{Khokhlov}},\ }\href@noop {} {\emph {\bibinfo {title} {Statistical {P}hysics
				of {M}acromolecules}}},\ AIP series in polymers and complex materials\
	(\bibinfo  {publisher} {AIP Press},\ \bibinfo {address} {New York},\ \bibinfo
	{year} {1994})\BibitemShut {NoStop}%
	\bibitem [{\citenamefont {Tracy}\ and\ \citenamefont
		{Widom}(1997)}]{tracy1997exact}%
	\BibitemOpen
	\bibfield  {author} {\bibinfo {author} {\bibfnamefont {C.~A.}\ \bibnamefont
			{Tracy}}\ and\ \bibinfo {author} {\bibfnamefont {H.}~\bibnamefont {Widom}},\
	}\bibfield  {title} {\enquote {\bibinfo {title} {On exact solutions to the
				cylindrical {P}oisson-{B}oltzmann equation with applications to
				polyelectrolytes},}\ }\href {\doibase 10.1016/S0378-4371(97)00229-X}
	{\bibfield  {journal} {\bibinfo  {journal} {Physica A: Statistical Mechanics
				and its Applications}\ }\textbf {\bibinfo {volume} {244}},\ \bibinfo {pages}
		{402} (\bibinfo {year} {1997})}\BibitemShut {NoStop}%
	\bibitem [{\citenamefont {T{\'e}llez}\ and\ \citenamefont
		{Trizac}(2006)}]{tellez.g:2006}%
	\BibitemOpen
	\bibfield  {author} {\bibinfo {author} {\bibfnamefont {G.}~\bibnamefont
			{T{\'e}llez}}\ and\ \bibinfo {author} {\bibfnamefont {E.}~\bibnamefont
			{Trizac}},\ }\bibfield  {title} {\enquote {\bibinfo {title} {Exact asymptotic
				expansions for the cylindrical {P}oisson--{B}oltzmann equation},}\
	}\href@noop {} {\bibfield  {journal} {\bibinfo  {journal} {Journal Stat.
				Mech.: Theory and Experiment}\ }\textbf {\bibinfo {volume} {2006}},\ \bibinfo
		{pages} {P06018} (\bibinfo {year} {2006})}\BibitemShut {NoStop}%
	\bibitem [{\citenamefont {Grosberg}\ \emph {et~al.}(2002)\citenamefont
		{Grosberg}, \citenamefont {Nguyen},\ and\ \citenamefont
		{Shklovskii}}]{grosberg.ayu:2002}%
	\BibitemOpen
	\bibfield  {author} {\bibinfo {author} {\bibfnamefont {A.~Y.}\ \bibnamefont
			{Grosberg}}, \bibinfo {author} {\bibfnamefont {T.~T.}\ \bibnamefont
			{Nguyen}}, \ and\ \bibinfo {author} {\bibfnamefont {B.~I.}\ \bibnamefont
			{Shklovskii}},\ }\bibfield  {title} {\enquote {\bibinfo {title} {Colloquium:
				{T}he physics of charge inversion in chemical and biological systems},}\
	}\href@noop {} {\bibfield  {journal} {\bibinfo  {journal} {Rev. Mod. Phys.}\
		}\textbf {\bibinfo {volume} {74}},\ \bibinfo {pages} {329} (\bibinfo {year}
		{2002})}\BibitemShut {NoStop}%
	\bibitem [{\citenamefont {Levin}(2002)}]{levin.y:2002}%
	\BibitemOpen
	\bibfield  {author} {\bibinfo {author} {\bibfnamefont {Y.}~\bibnamefont
			{Levin}},\ }\bibfield  {title} {\enquote {\bibinfo {title} {Electrostatic
				correlations: from plasma to biology},}\ }\href@noop {} {\bibfield  {journal}
		{\bibinfo  {journal} {Rep. Prog. Phys.}\ }\textbf {\bibinfo {volume} {65}},\
		\bibinfo {pages} {1577} (\bibinfo {year} {2002})}\BibitemShut {NoStop}%
	\bibitem [{\citenamefont {Vink}(1990)}]{vink.h:1990}%
	\BibitemOpen
	\bibfield  {author} {\bibinfo {author} {\bibfnamefont {H.}~\bibnamefont
			{Vink}},\ }\bibfield  {title} {\enquote {\bibinfo {title} {Ion binding in
				polyelectrolyte solutions and the {P}oisson--{B}oltzmann equation},}\
	}\href@noop {} {\bibfield  {journal} {\bibinfo  {journal} {J. Chem. Soc.,
				Faraday Trans.}\ }\textbf {\bibinfo {volume} {86}},\ \bibinfo {pages} {2607}
		(\bibinfo {year} {1990})}\BibitemShut {NoStop}%
	\bibitem [{\citenamefont {Fixman}(1979)}]{fixman.m:1979}%
	\BibitemOpen
	\bibfield  {author} {\bibinfo {author} {\bibfnamefont {M.}~\bibnamefont
			{Fixman}},\ }\bibfield  {title} {\enquote {\bibinfo {title} {The
				{P}oisson--{B}oltzmann equation and its application to polyelectrolytes},}\
	}\href@noop {} {\bibfield  {journal} {\bibinfo  {journal} {J. Chem. Phys.}\
		}\textbf {\bibinfo {volume} {70}},\ \bibinfo {pages} {4995} (\bibinfo {year}
		{1979})}\BibitemShut {NoStop}%
	\bibitem [{\citenamefont {Bocquet}\ \emph {et~al.}(2002)\citenamefont
		{Bocquet}, \citenamefont {Trizac},\ and\ \citenamefont
		{Aubouy}}]{bocquet.l:2002}%
	\BibitemOpen
	\bibfield  {author} {\bibinfo {author} {\bibfnamefont {L.}~\bibnamefont
			{Bocquet}}, \bibinfo {author} {\bibfnamefont {E.}~\bibnamefont {Trizac}}, \
		and\ \bibinfo {author} {\bibfnamefont {M.}~\bibnamefont {Aubouy}},\
	}\bibfield  {title} {\enquote {\bibinfo {title} {Effective charge saturation
				in colloidal suspensions},}\ }\href@noop {} {\bibfield  {journal} {\bibinfo
			{journal} {J. Chem. Phys.}\ }\textbf {\bibinfo {volume} {117}},\ \bibinfo
		{pages} {8138} (\bibinfo {year} {2002})}\BibitemShut {NoStop}%
	\bibitem [{\citenamefont {Deserno}\ \emph {et~al.}(2000)\citenamefont
		{Deserno}, \citenamefont {Holm},\ and\ \citenamefont {May}}]{deserno.m:2000}%
	\BibitemOpen
	\bibfield  {author} {\bibinfo {author} {\bibfnamefont {M.}~\bibnamefont
			{Deserno}}, \bibinfo {author} {\bibfnamefont {C.}~\bibnamefont {Holm}}, \
		and\ \bibinfo {author} {\bibfnamefont {S.}~\bibnamefont {May}},\ }\bibfield
	{title} {\enquote {\bibinfo {title} {Fraction of condensed counterions around
				a charged rod: Comparison of {P}oisson--{B}oltzmann theory and computer
				simulations},}\ }\href@noop {} {\bibfield  {journal} {\bibinfo  {journal}
			{Macromolecules}\ }\textbf {\bibinfo {volume} {33}},\ \bibinfo {pages} {199}
		(\bibinfo {year} {2000})}\BibitemShut {NoStop}%
	\bibitem [{\citenamefont {Stern}(1924)}]{stern.o:1924}%
	\BibitemOpen
	\bibfield  {author} {\bibinfo {author} {\bibfnamefont {O.}~\bibnamefont
			{Stern}},\ }\bibfield  {title} {\enquote {\bibinfo {title} {Zur {T}heorie der
				elektrolytischen {D}oppelschicht},}\ }\href@noop {} {\bibfield  {journal}
		{\bibinfo  {journal} {Z. Elektrochem}\ }\textbf {\bibinfo {volume} {30}},\
		\bibinfo {pages} {1014} (\bibinfo {year} {1924})}\BibitemShut {NoStop}%
	\bibitem [{\citenamefont {Bjerrum}(1926)}]{bjerrum.n:1926}%
	\BibitemOpen
	\bibfield  {author} {\bibinfo {author} {\bibfnamefont {N.}~\bibnamefont
			{Bjerrum}},\ }\bibfield  {title} {\enquote {\bibinfo {title}
			{{Untersuchungen} {\"u}ber {Ionenassoziation}. {I}. {Der} {Einfluss} der
				{Ionenassoziation} auf die {Aktivit{\"a}t} der {Ionen} bei m{\"a}ssigen
				{Assoziationsgraden}},}\ }\href@noop {} {\bibfield  {journal} {\bibinfo
			{journal} {Kgl. Dan. Vidensk. Selsk. Mat.-Fys. Medd.}\ }\textbf {\bibinfo
			{volume} {7}},\ \bibinfo {pages} {1} (\bibinfo {year} {1926})}\BibitemShut
	{NoStop}%
	\bibitem [{\citenamefont {Vinogradova}\ and\ \citenamefont
		{Silkina}(2025)}]{vinogradova.oi:2025}%
	\BibitemOpen
	\bibfield  {author} {\bibinfo {author} {\bibfnamefont {O.~I.}\ \bibnamefont
			{Vinogradova}}\ and\ \bibinfo {author} {\bibfnamefont {E.~F.}\ \bibnamefont
			{Silkina}},\ }\bibfield  {title} {\enquote {\bibinfo {title} {Conductivity of
				concentrated salt solutions},}\ }\href {\doibase 10.1063/5.0278320}
	{\bibfield  {journal} {\bibinfo  {journal} {J. Chem. Phys.}\ }\textbf
		{\bibinfo {volume} {163}},\ \bibinfo {pages} {044501} (\bibinfo {year}
		{2025})}\BibitemShut {NoStop}%
	\bibitem [{\citenamefont {Uematsu}\ \emph {et~al.}(2018)\citenamefont
		{Uematsu}, \citenamefont {Netz},\ and\ \citenamefont
		{Bonthuis}}]{uematsu.y:2018}%
	\BibitemOpen
	\bibfield  {author} {\bibinfo {author} {\bibfnamefont {Y.}~\bibnamefont
			{Uematsu}}, \bibinfo {author} {\bibfnamefont {R.~R.}\ \bibnamefont {Netz}}, \
		and\ \bibinfo {author} {\bibfnamefont {D.~J.}\ \bibnamefont {Bonthuis}},\
	}\bibfield  {title} {\enquote {\bibinfo {title} {Analytical interfacial layer
				model for the capacitance and electrokinetics of charged aqueous
				interfaces},}\ }\href@noop {} {\bibfield  {journal} {\bibinfo  {journal}
			{Langmuir}\ }\textbf {\bibinfo {volume} {34}},\ \bibinfo {pages} {9097}
		(\bibinfo {year} {2018})}\BibitemShut {NoStop}%
	\bibitem [{\citenamefont {Vinogradova}\ \emph {et~al.}(2024)\citenamefont
		{Vinogradova}, \citenamefont {Silkina},\ and\ \citenamefont
		{Asmolov}}]{vinogradova.oi:2024}%
	\BibitemOpen
	\bibfield  {author} {\bibinfo {author} {\bibfnamefont {O.~I.}\ \bibnamefont
			{Vinogradova}}, \bibinfo {author} {\bibfnamefont {E.~F.}\ \bibnamefont
			{Silkina}}, \ and\ \bibinfo {author} {\bibfnamefont {E.~S.}\ \bibnamefont
			{Asmolov}},\ }\bibfield  {title} {\enquote {\bibinfo {title} {Surface
				potentials of conductors in electrolyte solutions},}\ }\href@noop {}
	{\bibfield  {journal} {\bibinfo  {journal} {J. Chem. Phys.}\ }\textbf
		{\bibinfo {volume} {161}},\ \bibinfo {pages} {204702} (\bibinfo {year}
		{2024})}\BibitemShut {NoStop}%
	\bibitem [{\citenamefont {Tollmien}\ \emph {et~al.}(1961)\citenamefont
		{Tollmien}, \citenamefont {Schlichting}, \citenamefont {G{\"o}rtler},\ and\
		\citenamefont {Riegels}}]{Tollmien1961}%
	\BibitemOpen
	\bibfield  {author} {\bibinfo {author} {\bibfnamefont {W.}~\bibnamefont
			{Tollmien}}, \bibinfo {author} {\bibfnamefont {H.}~\bibnamefont
			{Schlichting}}, \bibinfo {author} {\bibfnamefont {H.}~\bibnamefont
			{G{\"o}rtler}}, \ and\ \bibinfo {author} {\bibfnamefont {F.~W.}\ \bibnamefont
			{Riegels}},\ }\enquote {\bibinfo {title} {{\"U}ber {F}l{\"u}ssigkeitsbewegung
			bei sehr kleiner {R}eibung},}\ in\ \href {\doibase
		10.1007/978-3-662-11836-8_43} {\emph {\bibinfo {booktitle} {Ludwig Prandtl
				Gesammelte Abhandlungen: zur angewandten Mechanik, Hydro- und
				Aerodynamik}}},\ \bibinfo {editor} {edited by\ \bibinfo {editor}
		{\bibfnamefont {F.~W.}\ \bibnamefont {Riegels}}}\ (\bibinfo  {publisher}
	{Springer Berlin Heidelberg},\ \bibinfo {address} {Berlin, Heidelberg},\
	\bibinfo {year} {1961})\ pp.\ \bibinfo {pages} {575--584}\BibitemShut
	{NoStop}%
	\bibitem [{\citenamefont {Vinogradova}\ and\ \citenamefont
		{Silkina}(2026)}]{vinogradova.sph:2026}%
	\BibitemOpen
	\bibfield  {author} {\bibinfo {author} {\bibfnamefont {O.~I.}\ \bibnamefont
			{Vinogradova}}\ and\ \bibinfo {author} {\bibfnamefont {E.~F.}\ \bibnamefont
			{Silkina}},\ }\bibfield  {title} {\enquote {\bibinfo {title} {Surface
				charge--potential relation for spherical particles in electrolyte
				solutions},}\ }\href@noop {} {\bibfield  {journal} {\bibinfo  {journal}
			{arXiv preprint arXiv:2610.03613}\ } (\bibinfo {year} {2026})}\BibitemShut
	{NoStop}%
	\bibitem [{\citenamefont {Bader}\ and\ \citenamefont
		{Ascher}(1987)}]{bader.g:1987}%
	\BibitemOpen
	\bibfield  {author} {\bibinfo {author} {\bibfnamefont {G.}~\bibnamefont
			{Bader}}\ and\ \bibinfo {author} {\bibfnamefont {U.}~\bibnamefont {Ascher}},\
	}\bibfield  {title} {\enquote {\bibinfo {title} {A new basis implementation
				for a mixed order boundary value {O}{D}{E} solver},}\ }\href@noop {}
	{\bibfield  {journal} {\bibinfo  {journal} {SIAM J. Sci. Comput.}\ }\textbf
		{\bibinfo {volume} {8}},\ \bibinfo {pages} {483} (\bibinfo {year}
		{1987})}\BibitemShut {NoStop}%
	\bibitem [{\citenamefont {Asmolov}\ \emph {et~al.}(2025)\citenamefont
		{Asmolov}, \citenamefont {Silkina},\ and\ \citenamefont
		{Vinogradova}}]{asmolov.es:2025b}%
	\BibitemOpen
	\bibfield  {author} {\bibinfo {author} {\bibfnamefont {E.~S.}\ \bibnamefont
			{Asmolov}}, \bibinfo {author} {\bibfnamefont {E.~F.}\ \bibnamefont
			{Silkina}}, \ and\ \bibinfo {author} {\bibfnamefont {O.~I.}\ \bibnamefont
			{Vinogradova}},\ }\bibfield  {title} {\enquote {\bibinfo {title} {Enhanced
				zeta potentials caused by surface ion mobilities},}\ }\href@noop {}
	{\bibfield  {journal} {\bibinfo  {journal} {Phys. Fluids}\ }\textbf {\bibinfo
			{volume} {37}},\ \bibinfo {pages} {082118} (\bibinfo {year}
		{2025})}\BibitemShut {NoStop}%
	\bibitem [{\citenamefont {Skolnick}\ and\ \citenamefont
		{Fixman}(1977)}]{skolnick.pe:1977}%
	\BibitemOpen
	\bibfield  {author} {\bibinfo {author} {\bibfnamefont {J.}~\bibnamefont
			{Skolnick}}\ and\ \bibinfo {author} {\bibfnamefont {M.}~\bibnamefont
			{Fixman}},\ }\bibfield  {title} {\enquote {\bibinfo {title} {Electrostatic
				persistence length of a wormlike polyelectrolyte},}\ }\href@noop {}
	{\bibfield  {journal} {\bibinfo  {journal} {Macromolecules}\ }\textbf
		{\bibinfo {volume} {10}},\ \bibinfo {pages} {944} (\bibinfo {year}
		{1977})}\BibitemShut {NoStop}%
	\bibitem [{\citenamefont {Khokhlov}\ and\ \citenamefont
		{Khachaturian}(1982)}]{khokhlov.pe:1982}%
	\BibitemOpen
	\bibfield  {author} {\bibinfo {author} {\bibfnamefont {A.~R.}\ \bibnamefont
			{Khokhlov}}\ and\ \bibinfo {author} {\bibfnamefont {K.~A.}\ \bibnamefont
			{Khachaturian}},\ }\bibfield  {title} {\enquote {\bibinfo {title} {On the
				theory of weakly charged polyelectrolytes},}\ }\href@noop {} {\bibfield
		{journal} {\bibinfo  {journal} {Polymer}\ }\textbf {\bibinfo {volume} {23}},\
		\bibinfo {pages} {1743} (\bibinfo {year} {1982})}\BibitemShut {NoStop}%
	\bibitem [{\citenamefont {Barrat}\ and\ \citenamefont
		{Joanny}(1993)}]{barrat.pe:1993}%
	\BibitemOpen
	\bibfield  {author} {\bibinfo {author} {\bibfnamefont {J.~L.}\ \bibnamefont
			{Barrat}}\ and\ \bibinfo {author} {\bibfnamefont {J.~F.}\ \bibnamefont
			{Joanny}},\ }\bibfield  {title} {\enquote {\bibinfo {title} {Persistence
				length of polyelectrolyte chains},}\ }\href@noop {} {\bibfield  {journal}
		{\bibinfo  {journal} {Europhysics Letters}\ }\textbf {\bibinfo {volume}
			{24}},\ \bibinfo {pages} {333} (\bibinfo {year} {1993})}\BibitemShut
	{NoStop}%
	\bibitem [{\citenamefont {Everaers}\ \emph {et~al.}(2002)\citenamefont
		{Everaers}, \citenamefont {Milchev},\ and\ \citenamefont
		{Yamakov}}]{everaers.r:2002}%
	\BibitemOpen
	\bibfield  {author} {\bibinfo {author} {\bibfnamefont {R.}~\bibnamefont
			{Everaers}}, \bibinfo {author} {\bibfnamefont {A.}~\bibnamefont {Milchev}}, \
		and\ \bibinfo {author} {\bibfnamefont {V.}~\bibnamefont {Yamakov}},\
	}\bibfield  {title} {\enquote {\bibinfo {title} {The electrostatic
				persistence length of polymers beyond the {OSF} limit},}\ }\href@noop {}
	{\bibfield  {journal} {\bibinfo  {journal} {The European Physical Journal E}\
		}\textbf {\bibinfo {volume} {8}},\ \bibinfo {pages} {3} (\bibinfo {year}
		{2002})}\BibitemShut {NoStop}%
	\bibitem [{\citenamefont {Rumyantsev}\ and\ \citenamefont
		{Johner}(2026)}]{rumyantsev.epl1:2026}%
	\BibitemOpen
	\bibfield  {author} {\bibinfo {author} {\bibfnamefont {A.~M.}\ \bibnamefont
			{Rumyantsev}}\ and\ \bibinfo {author} {\bibfnamefont {A.}~\bibnamefont
			{Johner}},\ }\bibfield  {title} {\enquote {\bibinfo {title} {Electrostatic
				persistence length revisited. {I}. {T}heory},}\ }\href@noop {} {\bibfield
		{journal} {\bibinfo  {journal} {arXiv preprint arXiv:2608.21617}\ } (\bibinfo
		{year} {2026})}\BibitemShut {NoStop}%
	\bibitem [{\citenamefont {Gavrilov}\ \emph {et~al.}(2026)\citenamefont
		{Gavrilov}, \citenamefont {Johner},\ and\ \citenamefont
		{Rumyantsev}}]{gavrilov.epl2:2026}%
	\BibitemOpen
	\bibfield  {author} {\bibinfo {author} {\bibfnamefont {A.~A.}\ \bibnamefont
			{Gavrilov}}, \bibinfo {author} {\bibfnamefont {A.}~\bibnamefont {Johner}}, \
		and\ \bibinfo {author} {\bibfnamefont {A.~M.}\ \bibnamefont {Rumyantsev}},\
	}\bibfield  {title} {\enquote {\bibinfo {title} {Electrostatic persistence
				length revisited. {II}. {S}imulations and comparison to experiment},}\
	}\href@noop {} {\bibfield  {journal} {\bibinfo  {journal} {arXiv preprint
				arXiv:2608.21627}\ } (\bibinfo {year} {2026})}\BibitemShut {NoStop}%
\end{thebibliography}
\end{document}